\documentclass[fleqn,usenatbib]{mnras}
\usepackage{newtxtext,newtxmath}

\usepackage[utf8]{inputenc}
\usepackage[T1]{fontenc}
\usepackage{ae,aecompl}
\usepackage{graphicx}	
\usepackage{amsmath}	
\usepackage{amssymb}	
\usepackage[normalem]{ulem}
\usepackage{multicol}
\usepackage{pdflscape}
\usepackage{lineno}

\newcommand{\noop}[1]{}

\usepackage{longtable}
\usepackage{array}
\usepackage{rotating}
\newcommand{\PreserveBackslash}[1]{\let\temp=\\#1\let\\=\temp}
\newcolumntype{C}[1]{>{\PreserveBackslash\centering}p{#1}}
\newcolumntype{R}[1]{>{\PreserveBackslash\raggedleft}p{#1}}
\newcolumntype{L}[1]{>{\PreserveBackslash\raggedright}p{#1}}

\newcommand{\mmode}[1]{\ifmmode{#1}\else{$#1$}\fi}

\newcommand{\tess}[0]{\emph{TESS}}

\newcommand{\Msolar}[0]{\mmode{\text{M}_{\odot}}}

\usepackage{CJKutf8}

\title[EHe TESS Photometry]{TESS Observations of Stochastic Low-frequency Variability in Extreme Helium Stars}
\author[C. L. Crawford et al.]{
Courtney L. Crawford$^{1}$\thanks{Email: courtney.crawford@sydney.edu.au},
C. Simon Jeffery $^{2}$,
May G. Pedersen$^{1}$,
Timothy R. Bedding$^{1}$,
\newauthor Benjamin T. Montet$^{3}$,
Geoffrey C. Clayton$^{4,5}$, 
Patrick Tisserand$^{6}$
\\
$^{1}$ Sydney Institute for Astronomy (SIfA), School of Physics, University of Sydney, NSW 2006, Australia\\
$^{2}$ Armagh Observatory and Planetarium, College Hill, Armagh BT61 9DG, United Kingdom\\
$^{3}$ School of Physics, University of New South Wales, Sydney, NSW 2052, Australia \\
$^{4}$ Space Science Institute (SSI), 4750 Walnut Street, Suite 205, Boulder, CO 80301, USA\\
$^{5}$ Department of Physics \& Astronomy, Louisiana State University, Baton Rouge, LA 70803, USA\\
$^{6}$ Sorbonne Universit\'e, CNRS, UMR 7095, Institut d'Astrophysique de Paris, 98 bis bd Arago, 75014 Paris, France\\
}

\date{Accepted xx. Received xx; in original form xx}
\pubyear{2025}
\begin{document}
\label{firstpage}
\pagerange{\pageref{firstpage}--\pageref{lastpage}}
\maketitle

\begin{abstract}

Extreme helium stars (EHes) are low-mass hydrogen-deficient stars thought to be the products of double white dwarf mergers. Despite prolonged ground-based observations, there is no consensus on the properties of their photometric variability. In this article, we present an analysis of TESS light curves for all known EHe stars, constituting the first population-level study of EHe photometric variability. We present updated TESS light curves for the two confirmed large-amplitude pulsators, V652 Her and BX Cir, and discuss the potential r-mode pulsators BD+37 442 and BD+37 1977. Notably, we found that the majority of EHe stars exhibit stochastic low-frequency (SLF) variability, or a signal with power increasing smoothly towards low frequencies, rather than peaks in the power spectrum corresponding to oscillation modes. We characterised the SLF variability of EHe stars using Gaussian process regression with a stochastically-driven/damped simple harmonic oscillator kernel and measured the characteristic timescale, low-frequency amplitude, and quality factor for each star. The variability timescales range from approximately 0.5 to 10 d and correlate strongly with stellar parameters such as density and radius. Further theoretical work is needed to determine the physical driving mechanism for the variability, but 1-D models of EHe stars suggest that thin subsurface convection zones may play a role.

\end{abstract}


\section{Introduction}
\label{sec:intro}

Extreme helium stars (EHes) are low-mass ($<$1\Msolar) giants and supergiants of spectral types A and B \citep{popper42,bidelman52,thackeray54}. 
Their surfaces show a highly processed chemical composition, with negligible or no hydrogen and overabundant carbon, nitrogen and, in some cases, oxygen \citep{jeffery11a}.
Their kinematics and Galactic distribution indicate roughly similar proportions of thin disk, thick disk, bulge and halo populations, despite their positions falling predominantly towards the Galactic disk and Galactic centre \citep{philipmonai24}. 
Distances measured by {\it Gaia} suggest they fall into two groups, with mean luminosities of around 1\,000 and 10\,000\,L$_{\odot}$, respectively. 

Following a proposal by \citet{webbink84}, \citet{saio00,saio02} argued that many observed properties of EHes could be accounted for by the product of merging two white dwarfs, including the absence of binary companions \citep{jeffery87}, masses and luminosities, surface compositions and total numbers of known EHes. 
Specifically, a merger of two helium white dwarfs (He+He) produces models consistent with observations of lower luminosity EHes, including the pulsating EHe star V652\,Her \citep{saio00}, whilst the merger of a helium white dwarf with a carbon-oxygen white dwarf (He+CO) produces models consistent with higher luminosity EHes such as PV\,Tel and V2244\,Oph \citep{saio02}. 
Subsequent studies supported this view \citep{clayton07,pandey06a,jeffery11a}.
The two EHe groups identified from distance-based luminosities  \citep{philipmonai24} correlate well with post-merger evolution tracks for He+He (low-luminosity) and He+CO (high luminosity), respectively \citep{zhang12a,zhang14,schwab18,schwab19,Crawford2024_dlhdcmodels}. 

However, the double white-dwarf theory relies heavily on EHe masses that are inferred from photometric variations assumed to be due to pulsations \citep{saio88b,saio95b,jeffery16a}. 
In only two stars have an unequivocal period measurement been established; V652\,Her and BX\,Cir are both pulsating stars with periods of $\approx 0.11$\,d \citep{landolt75,hill81,kilkenny95}. 
Both are members of the low-luminosity group and lie in the temperature domain where opacity from iron-group elements at $\approx 200\,000$\,K can drive large-amplitude pulsations \citep{saio93,saio95b}, similar to that which drives pulsations in $\beta$ Cepheids \citep{moskalik92}, hot subdwarfs \citep{charpinet97a}, and blue large-amplitude pulsators \citep{jeffery25}.

Early observations of more luminous EHes suggested several were variable \citep{landolt75}.
Possible periods were identified on the basis of a few pulsation cycles in FQ\,Aqr \citep{jeffery85a}, NO\,Ser \citep{jeffery86.hdef.b} and V2244 Oph \citep{morrison87a}, on timescales of 19--22, 5--8 and 10--11 days, respectively.  
In the cases of V2205\,Oph and V2076\,Oph, evidence for multi-periodic behaviour on timescales of 3--9 and 0.7--1.1\,d suggested the presence of pulsations \citep{jeffery85b,lynasgray87}. 
In addition to light variability, many EHes are also small-amplitude radial-velocity variables \citep{walker85,jeffery85b,jeffery87,jeffery92,lawson93}, with some evidence that light and velocity variations are correlated \citep{jeffery01c}.

More evidence for pulsations came from a correlation between the timescales of variation and the effective temperatures amongst EHe and related hydrogen-deficient stars \citep[][Fig. 2]{saio88b}, suggestive of the classical period mean-density relation for pulsating stars \citep{eddington17}. 
However, variability was not evidently ubiquitous: neither DN\,Leo nor V821\,Cen showed light variations in ground-based photometry \citep{hill84,grauer84,jeffery90} and extended ground-based campaigns failed to find unique stable pulsation periods in any EHe other than V652\,Her and BX\,Cir \citep{kilkenny99c,wright06}. 

Linear theory showed EHes to be pulsationally unstable in radial modes, provided the luminosity-to-mass ratio is sufficiently high \citep{saio88b,jeffery16a}. 
The pulsations are {\it strange} modes driven by the opacity mechanism. 
Non-linear pulsation models for EHe stars driven by Fe-bump opacities are well-behaved at luminosity-to-mass ratios $L/M \lesssim 3.3$ (solar units) \citep{jeffery22a,jeffery25} and provide excellent agreement with observations of the regular pulsators V652\,Her and BX\,Cir.
The high $L/M$ regime is less well explored by non-linear models \citep{montanes02.thesis}, and so EHe variability in general remains unexplained. 

The observations of variable EHes presented above, and before the era of photometry from space, were summarised by \citet{jeffery08.ibvs}. 
Broadly, they raised the following questions:  
\begin{enumerate}
\item Do all EHes vary in light? 
\item Are EHe variations periodic or stochastic? 
\item Can EHe variations be associated with a characteristic timescale? 
\item If so, are timescales correlated with other properties? 
\item Are EHe variations global or local? 
\item What are the driving mechanisms? 
\item Is there evidence for secular changes? 
\end{enumerate}
To improve on previous work, one must (a) acquire a substantially greater body of data, with better sampling and/or overall coverage and  (b) use analysis methods better suited to the nature of the phenomena. 

Regarding observations, since 2000 the advent of large-scale synoptic photometric surveys has provided substantial data over longer periods of time from the ground and, with greater precision, from space. 
Dedicated space photometry missions including {\it Kepler} \citep{Kepler} and {\it TESS} \citep{TESS} have radically changed our view of EHe variability. 
A {\it K2} lightcurve for the suspected EHe star HD~144941 showed evidence for surface spots \citep{jeffery18} and led to its reclassification as a helium-strong magnetic main-sequence star \citep{przybilla21a}.
A {\it K2} lightcurve for V348\,Sgr, which is simultaneously an active hot R Coronae Borealis (RCB) variable, a peculiar EHe, and the hydrogen-deficient central star of a planetary nebula \citep{pollacco90,leuenhagen94a,hecht98}, showed irregular small-amplitude variations at maximum light on time-scales expected for strange-mode pulsation in hot helium supergiants \citep{jeffery19a}. 
Early {\it TESS} targets included the two bright EHes: PV\,Tel \citep[Thackeray's star:][]{thackeray54} and V821\,Cen \citep[Popper's star:][]{popper42}. 
From single {\it TESS} sectors ($\approx 27$\,d), both showed irregular variations with amplitude $\approx$1\% on timescales of the order of a day \citep{jeffery20b}. 
Single-sector {\it TESS} lightcurves for two very hot EHes, BD+37\,1977 and BD+37\,442, showed multi-periodic variation, also with amplitude $\approx$1\%, that has been identified with possible $r$-mode oscillations having periods $\approx 0.2 - 1.1$\,d \citep{jeffery20c}. 
{\it TESS} has continued to observe fields containing EHes for many more sectors. 
These data have the potential to transform our understanding of EHe stars. 

In this article, we use the available {\it TESS} observations of all known EHe stars to perform the first population-level study of EHe variability and show for the first time that nearly all EHe stars exhibit stochastic low-frequency (SLF) variability. In Section~\ref{sec:data} we describe the \tess{} data available and our estimation of the contribution to the data from contamination. In Section~\ref{sec:vartypes} we summarise the variability in all EHe stars into a few broad types, with brief discussion of the previously known variables. In Section~\ref{sec:gpr} we discuss our characterisation of the SLF variability in EHe stars using Gaussian process regression (GPR) and the results of this analysis in Section~\ref{sec:gpr_results}. In Section~\ref{sec:causes} we discuss the possible origin of the SLF variability in EHe stars using theories from across the literature. We provide a brief comparison to other known SLF variables in Section~\ref{sec:slf_comparisons} and conclude with final discussions in Section~\ref{sec:conclusions}.


\section{TESS Data}
\label{sec:data}

As of the end of its second extended mission in late-2025, \tess{} has observed over 95\% of the sky, excluding only a small field towards the densest part of the Galactic Center. Sectors 91 and 92 were the first time \tess{} had observed many of the stars located towards the Galactic Center, where a significant portion of the EHe stars are located \citep{philipmonai24}. This means that 29 of the 30 known EHe stars have been observed, excluding only 2MASS~J18335703+0529170.

We used the Mikulski Archive for Space Telescopes (MAST) portal to download all publicly available light curves for our targets as of November 2025, listed in Table~\ref{tab:data}. We visually inspected all available light curves and manually chose the best one for our analysis, prioritising the longest baseline where possible and otherwise prioritising higher cadence data. For the majority of targets, we use the light curves created by the Quick Look Pipeline \citep[QLP,][]{QLP}, as they have the most sectors analysed. Note that these used the Full Frame Image (FFI) data for their analysis and they therefore have varying cadences, depending on when they were observed. If available, we used the SPOC 200-sec-cadence light curves \citep{SPOC}, as those have the highest precision and shortest cadence.

We ended up with \tess{} light curve data for 23 EHe stars. One star, EC~20236-5703, was present on the detector but does not have a light curve available due to it being an especially faint target ($T_{\rm mag}$=15.0136, $G$=14.5). Four additional stars had unusable light curves due to nearby brighter stars that dominated the flux--- these are V354~Nor, GALEX~J184559.8–413827, 2MASS~J18244794-2214291 and EC~20111-6902. One more star, V4732~Sgr, was removed from our analysis due to the field being too crowded to confidently isolate whether the observed non-periodic signal originates from our target. Finally, the QLP light curve for EC~19529-4430 was dominated by long term trends (likely instrumental), and we chose to use the \texttt{eleanor} \citep{eleanor} light curve for Sector 13 and discarded the later sectors.

\subsection{Contamination}
\label{subsec:contamination}

Because the pixel size of \tess{} is large (21\,arcsec) and many EHes lie in crowded regions of the sky, we paid extra attention to the possibility of contamination in the light curve. Contamination leads to dilution of the target's variability amplitudes and sometimes to spurious peaks in the power spectrum that are not intrinsic to the target (e.g. \citealt{Pedersen2023_contaminationFYPS}). To correct for amplitude dilution, we calculated the contamination factors for each target using the equation
\begin{equation}
    C = 1 - \frac{F^*_{\rm target}}{F^*_{\rm all}}.
    \label{eq:cont_fact}
\end{equation}
Here, $F^*_{\rm target}$ is the flux of the target inside the aperture mask retrieved from the light curve metadata\footnote{For SPOC light curves, it is straightforward using {\tt lightkurve} tools to extract the pipeline aperture. For QLP light curves, we retrieve the radius of the circle used to create the aperture mask using the column \texttt{BESTAP} and select all pixels that have pixel centres lying within that circle.} and $F^*_{\rm all}$ is the corresponding flux contribution inside the aperture from all nearby sources, including the target. If only the target contributes to the flux inside the aperture mask, then $C=0$. 
To estimate both $F^*_{\rm target}$ and $F^*_{\rm all}$, we simulated the \tess{} image by using the locations and brightnesses of all stars in the image reported by the TESS Input Catalogue (TIC). We estimated stellar flux values using $F^*_{\rm target} = 10^{-0.4T_{\rm mag}}$, where $T_{\rm mag}$ is the TESS magnitude, and then multiplying the target Pixel Response Function (PRF) obtained using the \texttt{TESS\_PRF} \texttt{python} package\footnote{\url{https://github.com/keatonb/TESS_PRF}}. Next, the flux inside the aperture mask from both the target and all stars was summed to give $F^*_{\rm target}$ and $F^*_{\rm all}$, respectively. The contamination factors are listed in Table~\ref{tab:data}.


\section{Characterising the Variability Types of EHes}
\label{sec:vartypes}

\begin{table*}
    \centering
    \caption{TESS data used in this work and variability classification}
    \label{tab:data}
    \begin{tabular}{lccccr}
        Star & Variability Type & LC Source & Sectors & TESS mag & $C$\\
        \hline
        NO Ser & SLF Variable & QLP & 80 & 9.718 & 0.334 \\
        V2244 Oph & SLF Variable & QLP & 80 & 10.591 & 0.313 \\
        PV Tel & SLF Variable & QLP & 13,66,93 & 9.239 & 0.048 \\
        LSS 99 & SLF Variable & QLP & 6,7,33,87 & 11.567 & 0.245 \\
        LSS 4357 & SLF Variable & QLP & 91,92 & 12.157 & 0.876 \\
        V1920 Cyg & SLF Variable & QLP & 41,54,55,74,75,81 & 10.218 & 0.372 \\
        CD-46 11775 & SLF Variable & QLP & 39,66,93 & 11.146 & 0.455 \\
        EC 19529-4430 & Non-Variable & QLP & 13,27,67,94 & 12.021 & 0.146 \\
        V2205 Oph & SLF Variable & QLP & 91 & 10.479 & 0.211 \\
        V5541 Sgr & SLF Variable & QLP & 92 & 13.004 & 0.810 \\
        V2076 Oph & SLF Variable & QLP & 92 & 9.683 & 0.381 \\
        BD+37 442 & r-mode & SPOC & 18,58 & 10.171 & 0.150 \\
        BD+37 1977 & r-mode & SPOC & 21,48 & 10.405 & 0.004 \\
        FQ Aqr & SLF Candidate & QLP & 55 & 9.341 & 0.146 \\
        V4732 Sgr & No LC - Crowded Field & QLP & 80,92 & 10.725 & 0.484 \\
        V354 Nor & No LC - Nearby Bright Star &   &   & 10.992 &   \\
        V821 Cen & SLF Variable & SPOC & 11,38,65 & 10.064 & 0.105 \\
        DN Leo & Non-Variable & SPOC & 45,46,72 & 10.136 & 0.002 \\
        V652 Her & Large Amplitude Pulsator & Custom (\texttt{eleanor}) & 79 & 10.744 &  0.015 \\
        BX Cir & Large Amplitude Pulsator & SPOC & 11,12,38,65 & 12.675 & 0.593 \\
        GALEX J184559.8–413827 & No LC - Nearby Bright Star &   &   & 14.728 &   \\
        EC 20236-5703 & No LC - faint &   &   & 15.014 &   \\
        LS IV+6 2 & Non-Variable & SPOC & 80 & 12.233 & 0.134 \\
        PG 1415+492 & Non-Variable & SPOC & 22,23,49,50 & 14.523 & 0.034 \\
        EC 20111-6902 & No LC - Nearby Bright Star &   &   & 15.496 &   \\
        EC 20262-6000 & Non-Variable & SPOC & 13,27,94 & 14.202 & 0.172 \\
        GALEX J191049.5-441713 & Non-Variable & SPOC & 13,27,67,94 & 13.070 & 0.311 \\
        2MASS J18244794-2214291 & No LC - Nearby Bright Star &   &   & 11.321 &   \\
        2MASS J18335703+0529170 & No LC - Not observed &   &   & 11.860 &   \\
        DY Cen & SLF Variable & QLP & 11,38,64,65 & 13.056 & 0.661 \\
    \end{tabular}
\end{table*}

\begin{figure}
    \centering
    \includegraphics[width=\columnwidth]{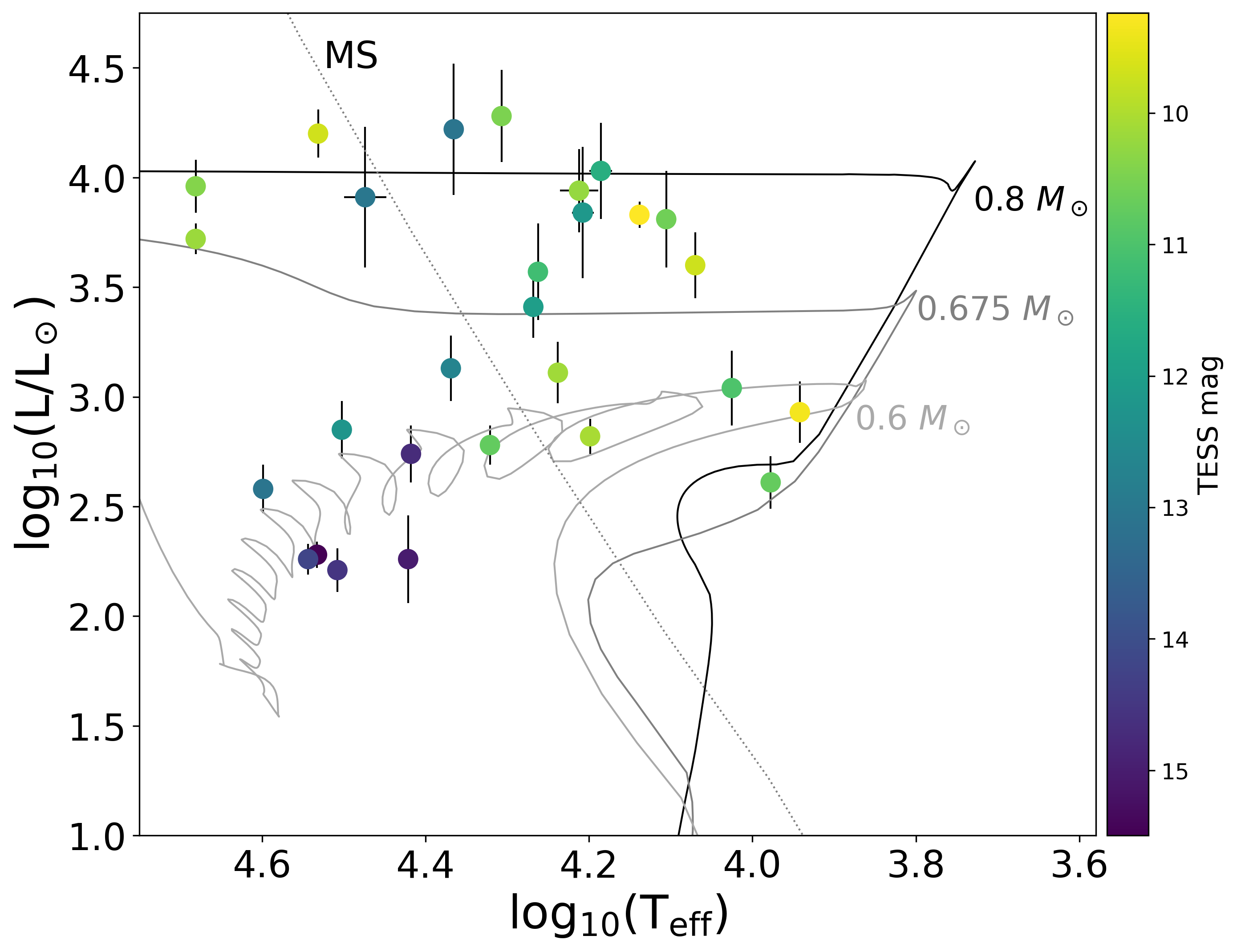}
    \includegraphics[width=\columnwidth]{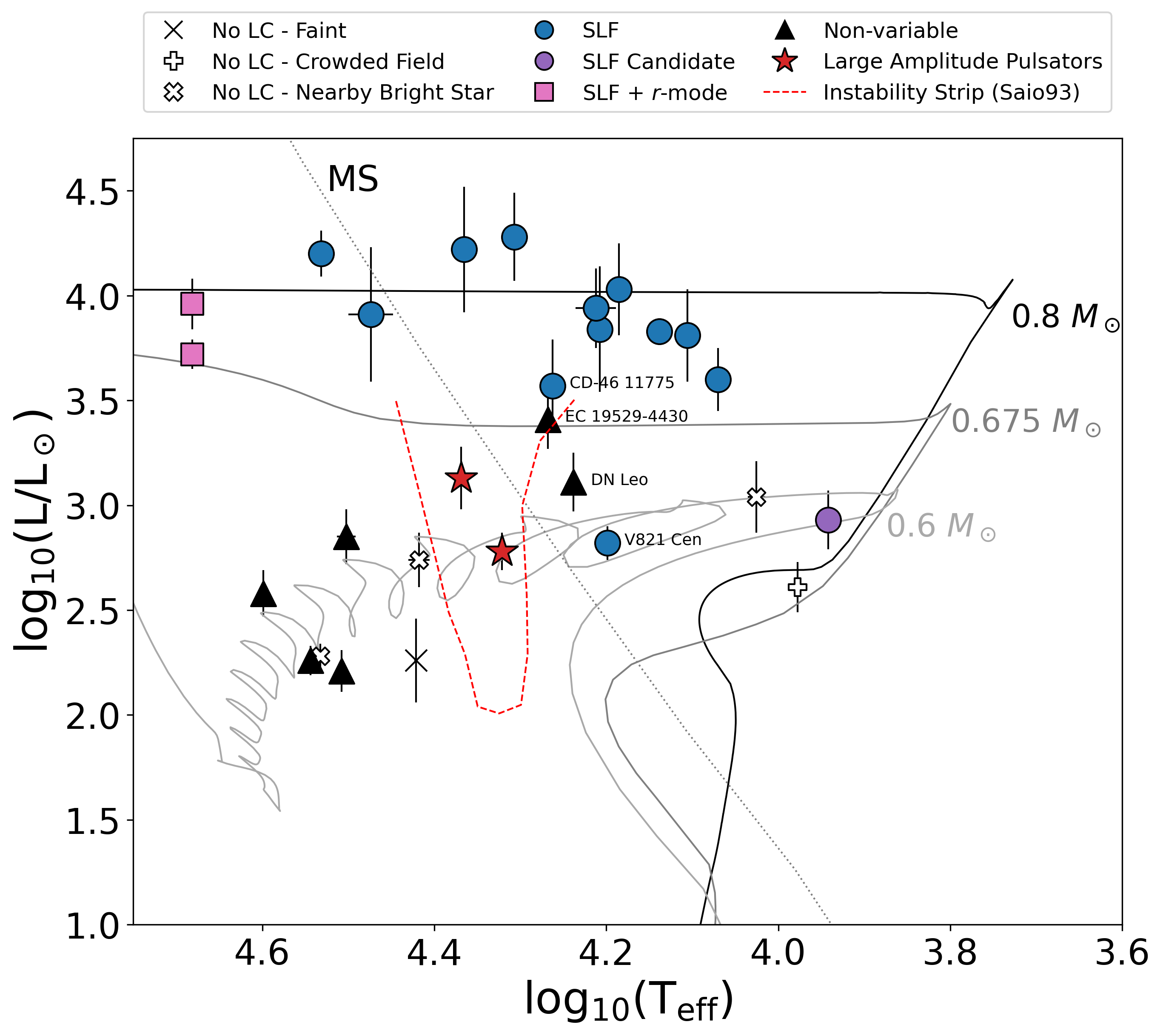}
    \caption{HRD of the EHe stars, with three evolutionary tracks of WD+WD mergers of varying total masses denoted with grey lines. Note that for the horizontal portion of the evolutionary track, the stars evolve leftwards, towards hotter temperatures as they contract. The track labelled 0.6 M$_\odot$ represents a He+He WD merger with mass ratio 1.0 \citep{zhang12a}, while the 0.675 M$_\odot$ and 0.8 M$_\odot$ tracks are CO+He mergers of the noted total mass with mass ratio of 0.675 and 0.45, respectively \citep{Crawford2024_dlhdcmodels}. Upper: The EHe stars are denoted in varying colours according to their brightness in the \tess{} filter. Lower: the symbols and markers denote the different classes of variables discussed in the text. The instability strip for the large amplitude pulsators \citep{saio93} is denoted in a dashed red line. 
    Note that the two EHe stars 2MASS~J18244794-2214291 and 2MASS~J18335703+0529170 do not appear as they do not have stellar parameters reported in \citet{philipmonai24} due to their recent discovery.}
    \label{fig:ehe_hrd}
\end{figure}

Our analysis of the available light curves for the EHe stars reveals a few broad categories of their variability: periodically variable, stochastically variable, and weakly- or non-variable, although the first two need not be mutually exclusive. The distribution of these variable types is shown in Figure~\ref{fig:ehe_hrd}. The periodic variable targets have all been previously studied. There are two sub-groups: the large amplitude pulsators\footnote{The name large amplitude pulsators is chosen to reflect their relationship to the blue large amplitude pulsators \citep{jeffery25}.} (BX\,Cir and V652\,Her) and the potential $r$-mode pulsators (BD+37~442 and BD+37~1977). The stochastically variable targets appear to belong to a growing class of so-called SLF variables that predominantly consist of massive and supergiant stars \citep[e.g.][]{Bowman2019_nature_slf}. Those in our sample include both pure SLF variables (with no oscillations) and the potential $r$-mode pulsators. Finally, the non-variable stars do not show significant evidence of oscillations or SLF variability.

\subsection{Periodic Variables}
\label{subsec:coherent}

\subsubsection{Large Amplitude Pulsators}
\label{subsubsec:non-linears}

The large amplitude pulsators---BX\,Cir and V652\,Her---are well-known and have been extensively studied from the ground since the 1970s \citep{landolt75,hill81,kilkenny95,Kilkenny2024_bxcir_v652her_tess}. These stars have significant, stable non-sinusoidal variations with amplitudes of $\sim$5\% in the visible and strong RV variations with semi-amplitudes up to 35 km/s. They have their own unique instability strip for H-deficient and solar-metallicity atmospheres \citep{saio93}. The exceptional stability of their pulsations permits precise measurement of period evolution, providing direct evidence for ongoing stellar contraction \citep{kilkenny82,kilkenny05}. Most recently, the \tess{} light curve of BX\,Cir in sectors 11, 12, and 38 was analysed by \citet{Kilkenny2024_bxcir_v652her_tess}, along with a recent South African Astronomical Observatory (SAAO) light curve of V652\,Her. In this section we provide an updated \tess{} light curve for BX\,Cir with a detailed contamination study and the first \tess{} light curve for V652\,Her. For both stars we also report the significant observable frequencies in the power spectrum.

TESS data for V652\,Her have not been previously published, but it has been observed several times from the ground \citep{landolt75,kilkenny82,kilkenny88,kilkenny91,Kilkenny96,kilkenny05,Kilkenny2024_bxcir_v652her_tess}. This star was observed in Sector 79. Despite being relatively bright ($V=10.51$) and isolated, it was not processed by SPOC. A QLP light curve is available but it suffers from a few instrumental artefacts. In addition, QLP light curves use the time stamp data from the \tess{} FFIs, which are not properly barycentric corrected for targets near to the edges of the detector, due to the differences in light-travel time for stars at the center of the detector compared to the edges \citep{TASOC_timestamps}. Therefore, the QLP time stamps are systematically offset by roughly 2.41 seconds on the edge where V652\,Her is located. Instead, we extracted a new light curve using \texttt{eleanor} \citep{eleanor}, which has a proper barycentric correction on the edges of the detector. This allowed us to recover a few more cycles of V652\,Her's oscillations than those available in the QLP light curve, and to compare with the ephemeris data published in \citet{Kilkenny2024_bxcir_v652her_tess} using exact time stamps. 

\begin{figure}
    \centering
    \includegraphics[width=\columnwidth]{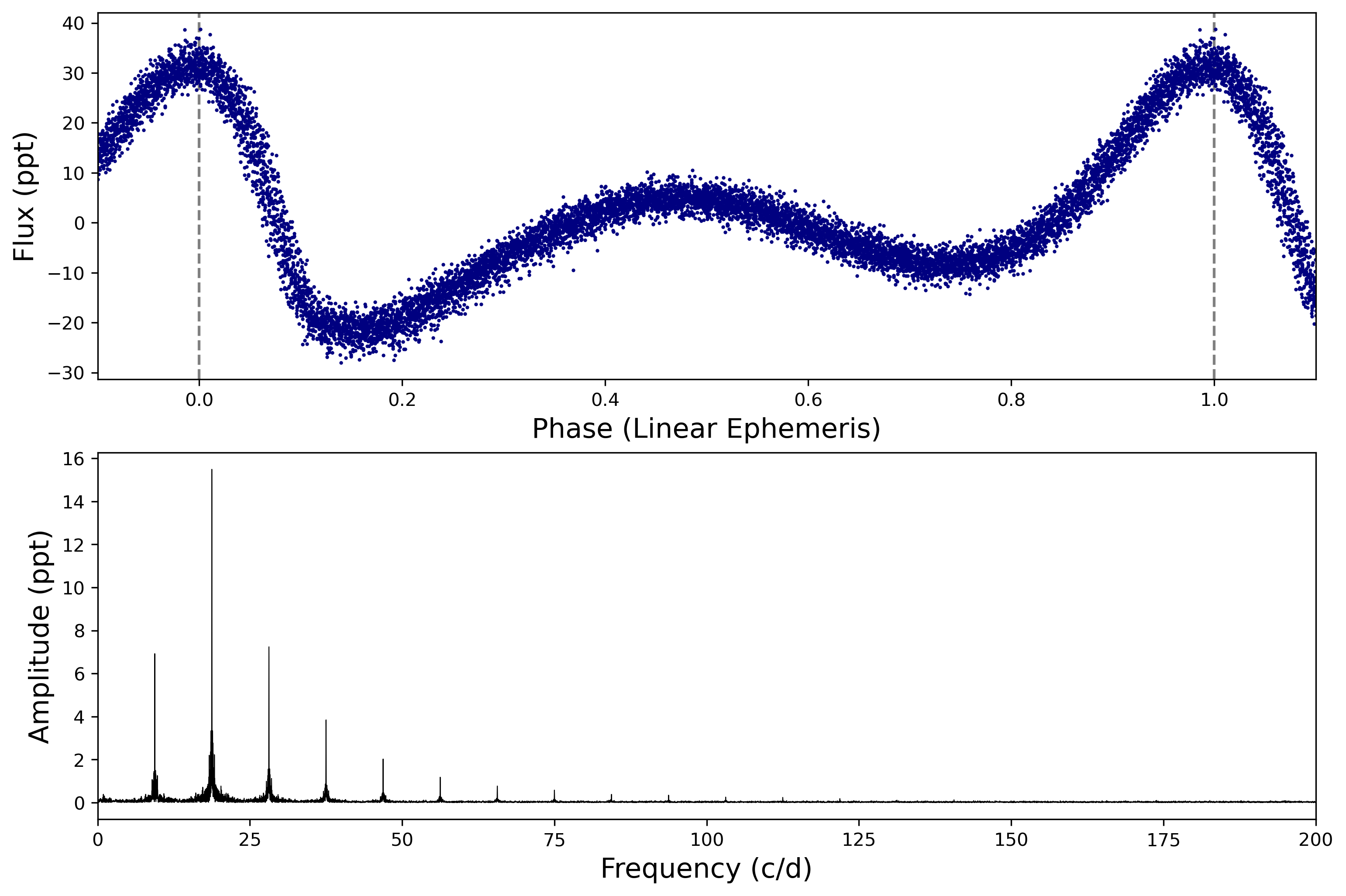}
    \caption{Upper: The \tess{} light curve of V652\,Her folded using a linear ephemeris with period calculated via iterative prewhitening (See Table~\ref{tab:v652her}). Vertical dashed grey lines indicate the beginning and end of the cycle.
    Lower: The amplitude spectrum of the \tess{} light curve.}
    \label{fig:v652her}
\end{figure}

The \tess{} light curve from {\tt eleanor} is presented in Figure~\ref{fig:v652her}, folded by the oscillation period (0.106690847 days). In the lower panel, we show the amplitude spectrum for this light curve, which exhibits the clear harmonic structure characteristic of these types of variables. In Table~\ref{tab:v652her} we list the 13 frequencies found via iterative pre-whitening, which constitute a harmonic series. We also note that the oscillation period we found from the \tess{} data (roughly 0.1067 days) is the shortest measured period for this star in the literature, and is consistent with the picture that this star has been spinning up over time as the star contracts. The first measured period for V652\,Her is 0.107995 days from \citet{landolt75}, indicating a change in period of 1.86 minutes in the past 50 years. \citet{Kilkenny2024_bxcir_v652her_tess} reported a quartic ephemeris based on 50 years of data prior to TESS. However, folding the \tess{} light curve of V652\,Her on this ephemeris does not produce a coherent phase curve, demonstrating that the published ephemeris is no longer valid at recent epochs. Calculation of a new ephemeris is left to later works because \tess{} is expected to observe V652\,Her again in 2027 during Sector 117.

\begin{table}
    \centering
    \caption{Frequencies prewhitened from the V652\,Her \tess{} light curve}
    \label{tab:v652her}
    \begin{tabular}{cccccc}
         Label & Frequency & $\sigma_{\rm freq}$ & Amplitude & Phase & S/N \\
          & (c/d) & (c/d) & (ppt) &  &  \\
         \hline
        1$f$ & 9.3729 & 0.00017 & 6.969 & 0.636 & 7.487 \\
        2$f$ & 18.7474 & 0.00008 & 15.524 & 0.000 & 7.841 \\
        3$f$ & 28.1213 & 0.00016 & 7.280 & 0.676 & 7.772 \\
        4$f$ & 37.4956 & 0.00030 & 3.872 & 0.161 & 7.659 \\
        5$f$ & 46.8700 & 0.00058 & 2.053 & 1.000 & 7.636 \\
        6$f$ & 56.2447 & 0.00098 & 1.206 & 0.999 & 7.665 \\
        7$f$ & 65.6183 & 0.00150 & 0.787 & 0.612 & 7.321 \\
        8$f$ & 74.9919 & 0.00201 & 0.587 & 0.098 & 7.389 \\
        9$f$ & 84.3672 & 0.00307 & 0.385 & 1.000 & 5.885 \\
        10$f$ & 93.7388 & 0.00334 & 0.353 & 0.559 & 6.154 \\
        11$f$ & 103.1142 & 0.00447 & 0.264 & 0.000 & 5.247 \\
        12$f$ & 112.4880 & 0.00486 & 0.243 & 0.658 & 5.090 \\
        13$f$ & 121.8632 & 0.00659 & 0.179 & 1.000 & 4.518 \\
    \end{tabular}
\end{table}

\begin{figure}
    \centering
    \includegraphics[width=\columnwidth]{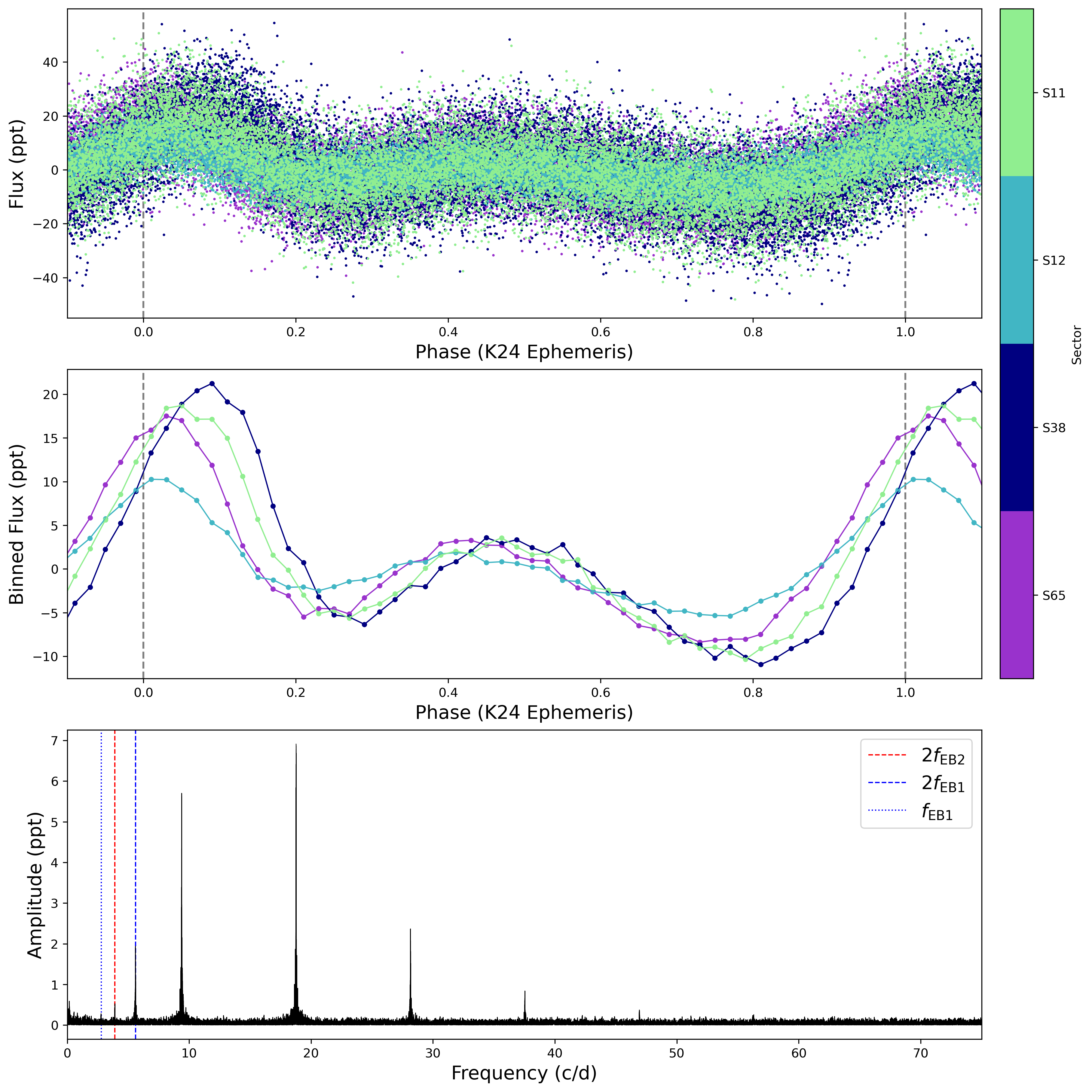}
    \caption{Upper: The \tess{} light curve of BX\,Cir folded using the cubic ephemeris from \citet{Kilkenny2024_bxcir_v652her_tess}. Different coloured points denote different \tess{} sectors as indicated in the colorbar. Vertical dashed grey lines indicate the beginning and end of the cycle.
    Middle: Same as the upper plot except the flux data has been binned to a width of 0.02 in phase to showcase the slight differences between sectors. Note that the y-axis limits are different to the upper plot.
    Lower: The amplitude spectrum of the full \tess{} light curve. We denote contaminating peaks from two nearby eclipsing binaries (EB) in blue and red vertical lines. Blue denotes the closer EB, and red the further EB (see Appendix~\ref{app:tess-localize}). Dashed vertical lines indicate the harmonic of the orbital frequency, and dotted lines indicate the orbital frequency (if visible). Note that $f_{\rm EB1}$, the orbital frequency for the nearer EB, does not appear in Table~\ref{tab:bxcir} because it is only visible in the amplitude spectrum of the full \tess{} light curve.}
    \label{fig:bxcir}
\end{figure}

BX\,Cir has been observed by \tess{} in sectors 11, 12, 38, and 65. The first three sectors were analysed by \citet{Kilkenny2024_bxcir_v652her_tess}. We show the SPOC 2-minute light curves\footnote{We note that SPOC light curves have a barycentre correction calculated according to the position of the star.} for all sectors in the upper panel of Figure~\ref{fig:bxcir}, phase folded using the cubic ephemeris measured by \citet{Kilkenny2024_bxcir_v652her_tess}, which was calculated using the timings from \tess{} Sectors 11, 12, and 38. In the middle panel we show the same light curves binned to a width of 0.02 in phase. In the lower panel we show the amplitude spectrum of the full \tess{} light curve. Using iterative prewhitening on each \tess{} sector separately, we extracted all frequencies with SNR $>$ 4 and list them in Table~\ref{tab:bxcir}. The prewhitening was done in order of highest to lowest SNR peaks in the Lomb-Scargle periodogram \citep{Lomb1976,Scargle1982} using the same procedure as in \cite{Pedersen2025_slf}. We report uncertainties on the frequencies calculated using Eqn~15 from \citet{Aerts2021_review}. 
All our extracted frequencies agree with those found by \citet{Kilkenny2024_bxcir_v652her_tess} to within the uncertainties. 

BX\,Cir has three spurious frequencies in its \tess{} power spectrum at $f_{\rm EB1}$=2.798 c/d, $f_{\rm EB2}$=3.910 c/d and $2f_{\rm EB1}$=5.597 c/d, as first noted by \citet{Kilkenny2024_bxcir_v652her_tess}, see the lower panel of Figure~\ref{fig:bxcir}. We note that $f_{\rm EB1}$=2.798 c/d is only detectable in the amplitude spectrum of the full \tess{} light curve, due its low signal to noise, and therefore does not appear in Table~\ref{tab:bxcir}. Using \texttt{tess\_localize} \citep{TESS-localize} we identified these two signals as contamination due to the known eclipsing binaries Gaia~DR3~5851089030362801152 (EB1, G=17.024, blue vertical lines) and Gaia~DR3~5851089339600429824 (EB2, G=17.215, red vertical line), which have orbital periods of 0.357381 days and 0.511548 days, respectively \citep{GaiaDR3_EBcatalog}. For EB1, the closer system, we observe both the orbital frequency and its harmonic (dotted and dashed lines, respectively), whereas for EB2 we observe only the harmonic of the orbital frequency. These frequencies are labelled as such in Table~\ref{tab:bxcir} for clarity.
Appendix~\ref{app:tess-localize} gives more details on the \texttt{tess\_localize} analysis. We emphasise the importance of contamination analyses; the more distant contaminating system, EB2, is roughly 77 arcsec ($\approx$ 3.5 \emph{TESS} pixels) away from BX\,Cir and does not lie within the SPOC aperture.

The oscillation period of BX\,Cir is changing over time at a measured rate of $dP/dn = -3.45\times 10^{-10}$ d, where $n$ counts the oscillation cycles \citep{Kilkenny2024_bxcir_v652her_tess}. Using this rate, we predict a change in the oscillation frequency over 5 years (roughly the baseline of the \tess{} observations of this star) to be $\Delta f = 5.21\times 10^{-4}$ c/d. This is on the same order of magnitude as the uncertainty on our measured frequencies, and therefore we do not robustly measure this frequency change. However, a slight frequency change can be inferred due to the misalignment of the light maxima between sectors. This can be seen in the folded and binned light curve shown in the middle panel of Figure~\ref{fig:bxcir}. In particular, the light maxima in Sector 38 appear slightly delayed compared to the first two sectors of \tess{} observations, whereas Sector 65 is more aligned with the first two sectors. BX\,Cir is planned to be reobserved by \tess{} during early 2026 and early 2027. After those data are released, one could potentially measure a new ephemeris and perform a direct measurement of the period contraction over the duration of the \tess{} mission (predicted $\Delta f = 8.3\times 10^{-4}$ c/d, $\Delta P=-0.81$ s).

\begin{table}
    \centering
    \caption{Frequencies prewhitened from the BX\,Cir \tess{} light curve}
    \label{tab:bxcir}
    \begin{tabular}{cccccc}
        Label & Frequency & $\sigma_{\rm freq}$ & Amplitude & Phase & S/N  \\
          & (c/d) & (c/d) & (ppt) &  &  \\
         \hline
        2$f_{\rm EB1}$ & 5.5972 & 0.00255 & 2.171 & 0.129 & 6.658 \\
        $f$ & 9.3859 & 0.00085 & 6.530 & 0.044 & 8.072 \\
        2$f$ & 18.7705 & 0.00064 & 8.669 & 0.037 & 8.364 \\
        3$f$ & 28.1564 & 0.00176 & 3.147 & 0.837 & 7.273 \\
        4$f$ & 37.5390 & 0.00528 & 1.048 & 0.881 & 4.518 \\
        \hline
        2$f_{\rm EB2}$ & 3.9098 & 0.00546 & 0.518 & 0.289 & 4.257 \\
        2$f_{\rm EB1}$ & 5.5961 & 0.00240 & 1.178 & 1.000 & 6.772 \\
        $f$ & 9.3846 & 0.00080 & 3.540 & 0.149 & 7.976 \\
        2$f$ & 18.7706 & 0.00065 & 4.334 & 0.804 & 8.077 \\
        3$f$ & 28.1562 & 0.00160 & 1.762 & 0.103 & 7.073 \\
        4$f$ & 37.5416 & 0.00457 & 0.618 & 0.587 & 5.440 \\
        \hline
        2$f_{\rm EB2}$ & 3.9048 & 0.00553 & 0.840 & 0.788 & 4.705 \\
        2$f_{\rm EB1}$ & 5.5962 & 0.00186 & 2.499 & 0.947 & 8.520 \\
        $f$ & 9.3853 & 0.00063 & 7.322 & 0.914 & 9.932 \\
        2$f$ & 18.7706 & 0.00050 & 9.342 & 0.000 & 9.988 \\
        3$f$ & 28.1563 & 0.00123 & 3.764 & 0.621 & 8.880 \\
        4$f$ & 37.5398 & 0.00402 & 1.156 & 0.782 & 5.655 \\
        \hline
        2$f_{\rm EB2}$ & 3.9099 & 0.00584 & 0.593 & 0.632 & 4.229 \\
        2$f_{\rm EB1}$ & 5.5974 & 0.00182 & 1.901 & 1.000 & 8.726 \\
        $f$ & 9.3857 & 0.00061 & 5.673 & 0.012 & 10.960 \\
        2$f$ & 18.7716 & 0.00044 & 7.886 & 0.008 & 11.262 \\
        3$f$ & 28.1563 & 0.00113 & 3.069 & 0.000 & 9.645 \\
        4$f$ & 37.5432 & 0.00349 & 0.991 & 0.983 & 6.753 \\
    \end{tabular}
\end{table}

\subsubsection{Potential $r$-mode Pulsators}
\label{subsubsec:rmodes}

BD+37~442 and BD+37~1977 were identified to have small amplitude peaks in their amplitude spectra by \citet{jeffery20c} using the SPOC 2-minute light curves from \tess{} Sectors~18 and 21, respectively. They discussed a few possible origins for these signals such as binarity and rotation, however their favoured explanation for was surface Rossby waves ($r$-modes). We found BD+37~442 to have signals at $\sim$1.78 and $\sim$3.55 c/d, where the latter signal is the harmonic of the former (frequency ratio = 1.999). Similarly, we find BD+37~1977 to have signals at $\sim$0.87 and $\sim$1.73 c/d, which may represent a harmonic sequence, however, the frequency ratio is 1.977, consistent with the findings of \citet{jeffery20c}. Since the low frequency peak is low signal-to-noise, we may be detecting an alias of the true frequency.

\begin{figure}
    \centering
    \includegraphics[width=\columnwidth]{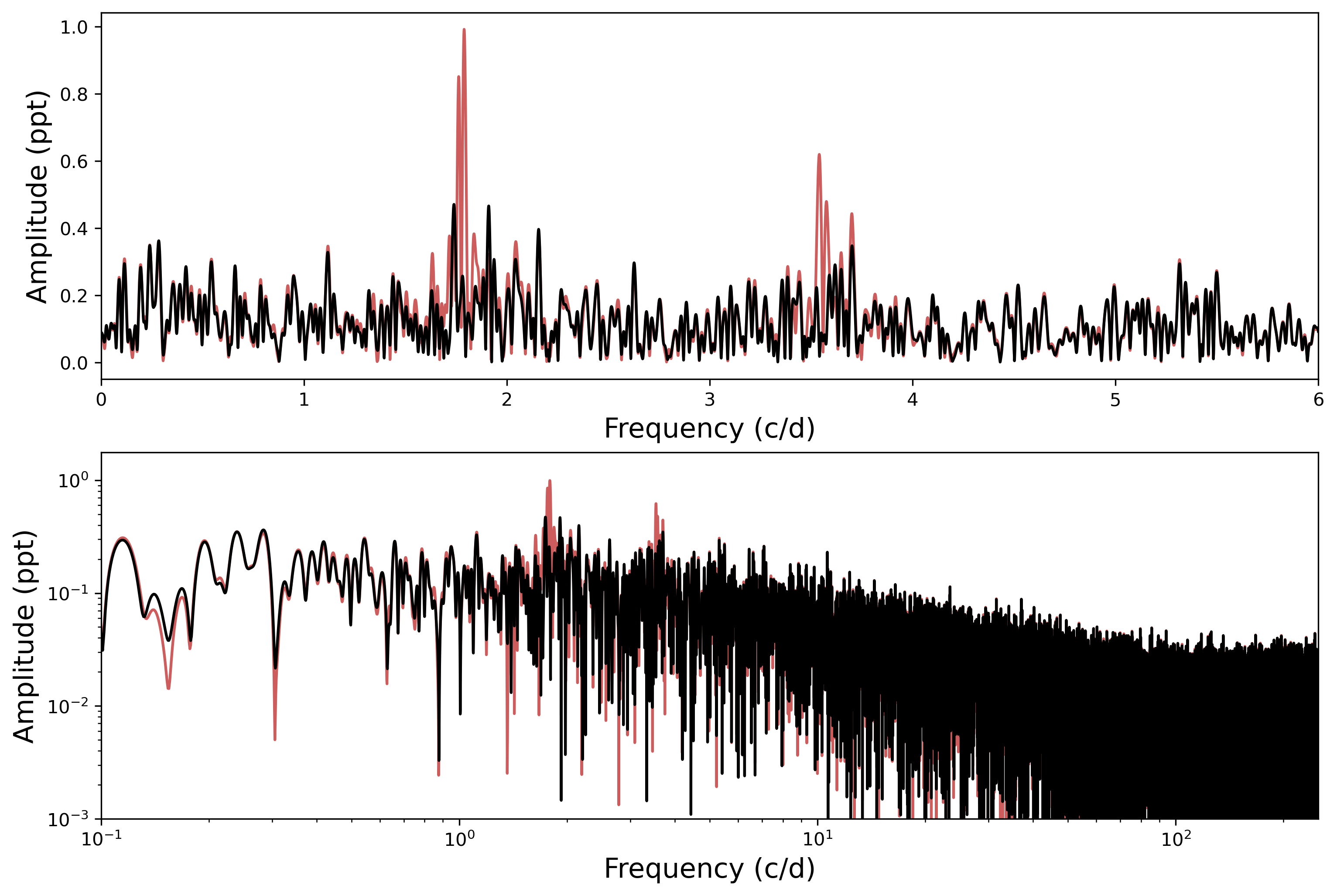}
    \caption{Amplitude spectrum of BD+37~442 before and after iterative prewhitening (red and black, respectively). Upper: Frequency and amplitude are in linear space and showcase the lowest frequencies to highlight the modes. Lower: Frequency and amplitude are in logarithmic space and showcase the full spectral range, highlighting the SLF variability signal.}
    \label{fig:bd+37_442}
\end{figure}

\begin{figure}
    \centering
    \includegraphics[width=\columnwidth]{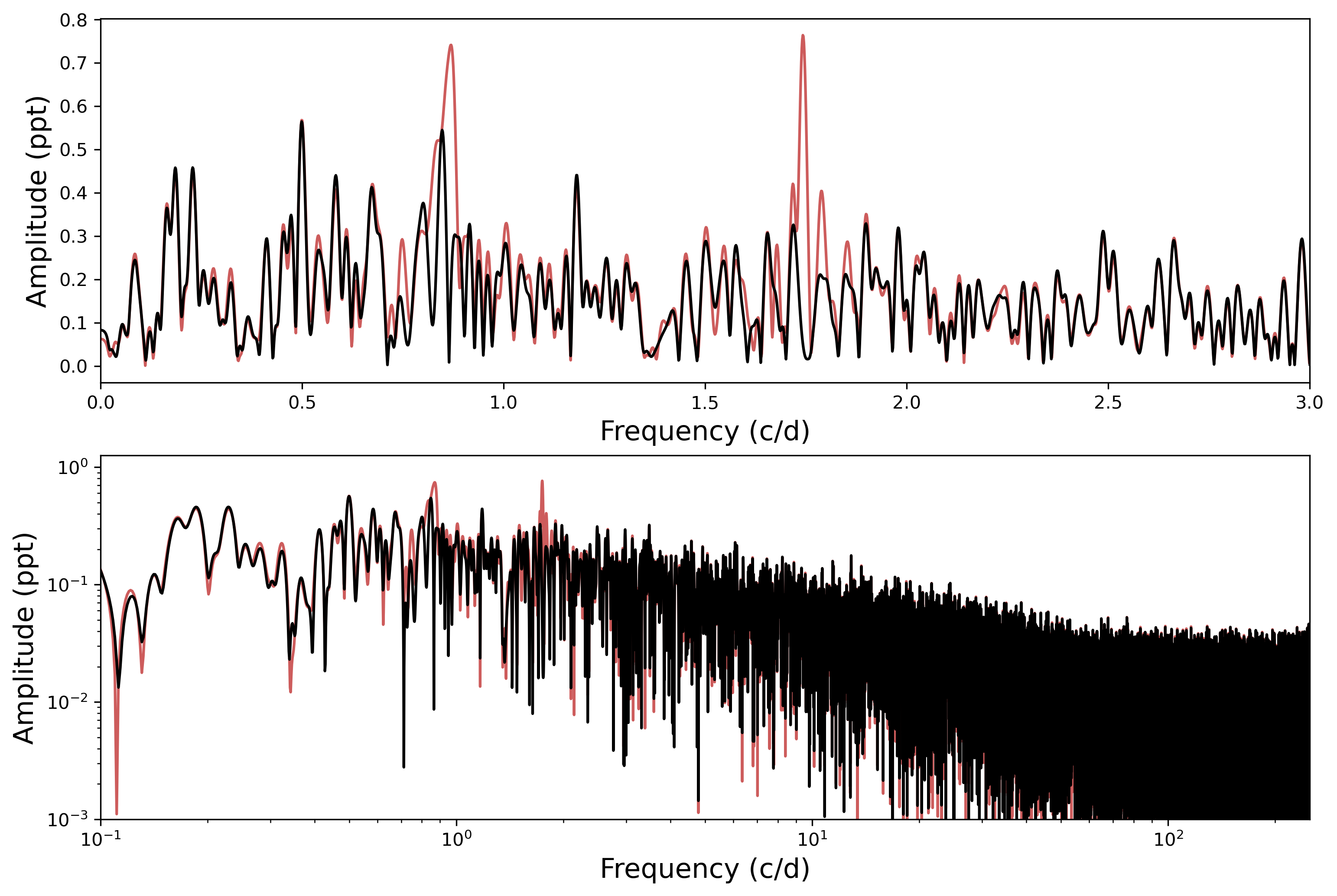}
    \caption{Amplitude spectrum of BD+37~1977 before and after iterative prewhitening (red and black, respectively). Upper: Frequency and amplitude are in linear space and showcase the lowest frequencies to highlight the modes. Lower: Frequency and amplitude are in logarithmic space and showcase the full spectral range, highlighting the SLF variability signal.}
    \label{fig:bd+37_1977}
\end{figure}

Here, we present the most recent \tess{} light curves for these targets.
BD+37~442 was observed in Sectors 18, 58, and 85. SPOC 2-minute light curves are available for Sectors 18 and 58, and a QLP light curve is available for Sector 85. A SPOC 20-second light curve is available for Sector 58. BD+37~1977 was observed in Sectors 21 and 48. A SPOC 2-minute light curve is also available for both sectors, and a SPOC 20-second light curve is available for Sector 48. To maintain consistency between the two stars while maximising the length of the observations, we use the SPOC 2-minute cadence light curves for all sectors, ignoring the noisier QLP light curve for Sector 85 of BD+37~442. We use iterative prewhitening to extract all frequencies with S/N $>$ 4 from the full light curves. These frequencies are listed in Tables~\ref{tab:bd+37_442} and~\ref{tab:bd+37_1977}, respectively. We show the amplitude spectra for the two stars before and after prewhitening in Figures~\ref{fig:bd+37_442} and~\ref{fig:bd+37_1977}. The frequencies extracted for these two stars agree within uncertainties with those found in \citet{jeffery20c}.

For both BD+37~442 and BD+37~1977, we found that after prewhitening the observed signals, there is a correlated low-frequency noise signal that persists in both amplitude spectra (See the lower panels of Figures~\ref{fig:bd+37_442} and~\ref{fig:bd+37_1977}). We identified this with the phenomenon of SLF variability which will be discussed further in Section~\ref{subsec:stochastic}. Additionally, we confirmed using \texttt{tess\_localize} \citep{TESS-localize} that the frequencies extracted from these light curves originate from the target stars.

\begin{table}
    \centering
    \caption{Frequencies prewhitened from the BD+37~442 \tess{} light curve}
    \label{tab:bd+37_442}
    \begin{tabular}{cccccc}
        Label & Frequency & $\sigma_{\rm freq}$ & Amplitude & Phase & S/N  \\
          & (c/d) & (c/d) & (ppt) &  &  \\
         \hline
        1$f$ & 1.7783 & 0.00067 & 1.186 & 0.613 & 6.307 \\
        2$f$ & 3.5550 & 0.00103 & 0.764 & 0.256 & 5.174 \\
    \end{tabular}
\end{table}

\begin{table}
    \centering
    \caption{Frequencies prewhitened from the BD+37~1977 \tess{} light curve}
    \label{tab:bd+37_1977}
    \begin{tabular}{cccccc}
         Label & Frequency & $\sigma_{\rm freq}$ & Amplitude & Phase & S/N  \\
          & (c/d) & (c/d) & (ppt) &  &  \\
         \hline
        1$f$ & 0.8782 & 0.00090 & 0.927 & 0.406 & 4.090 \\
        2$f$ & 1.7370 & 0.00103 & 0.808 & 0.890 & 4.616 \\
    \end{tabular}
\end{table}

\subsection{Stochastic Variables}
\label{subsec:stochastic}

The majority of EHe light curves from \tess{} indicate the presence of SLF variability. One candidate SLF variable, FQ~Aqr, has clear variability but with a timescale comparable to the observation time of a single \tess{} sector, making it difficult to characterise. SLF variability refers to a stochastic signal with Fourier power increasing towards the lowest frequencies. This type of signal may be referred to as "red noise", indicating its power at low frequencies, "Brown noise" (a reference to Brownian motion), or "correlated noise", since each measurement in the time series is dependent on the previous one. In terms of their statistical properties, these descriptors indicate similar signals, with only minor differences in the power law dependence of the power density on the frequency. We fit and characterise the SLF variability of the EHe stars in Section~\ref{sec:gpr}.

One SLF variable in our sample, LSS~4357, shows an additional peak in its power spectrum at approximately 3.3 c/d, which we identify as originating from the nearby RR Lyrae star OGLE-BLG-RRLYR-1881, located roughly 14" away and with a known pulsation period of 0.305676 days \citep{OGLE_BLG_RRL}. Since the contaminating star lies within 0.5 pixels of the target, we do not use \texttt{tess\_localize} to confirm the location of this signal. We note that this star also has many nearby eclipsing binaries (see the high contamination factor in Table~\ref{tab:data}) but there are no other spurious peaks visible in the power spectrum.

\subsection{Non-Variables}
\label{subsec:weak}

For many of the faintest EHe stars, the light curves show no conclusive evidence of variability, although there is often weak, low signal-to-noise (S/N $<$ 25, note the high S/N of the SLF variables in Table~\ref{tab:gpr_fits}) evidence of correlated signals. Due to the weakness of this evidence, we cannot rule out the possibility of extrinsic instrumental noise in the light curves. We denote six stars as non-detections of variability: LS~IV+6~2, PG~1415+492, EC~20262-6000, GALEX~J191049.5-441713, DN~Leo, and EC~19529-4430. However, we note that neither DN~Leo ($T_{\rm mag}$ = 10.14) nor EC~19529-4430 ($T_{\rm mag}$ = 12.02) are particularly faint relative to the full sample, and therefore indicate a true lack of variability for these two stars. Recall that EC~19529-4430 additionally exhibited strong instrumental variation in its QLP light curve, hence our preference for the \texttt{eleanor} light curve. 


\begin{figure*}
    \centering
    \includegraphics[width=\textwidth]{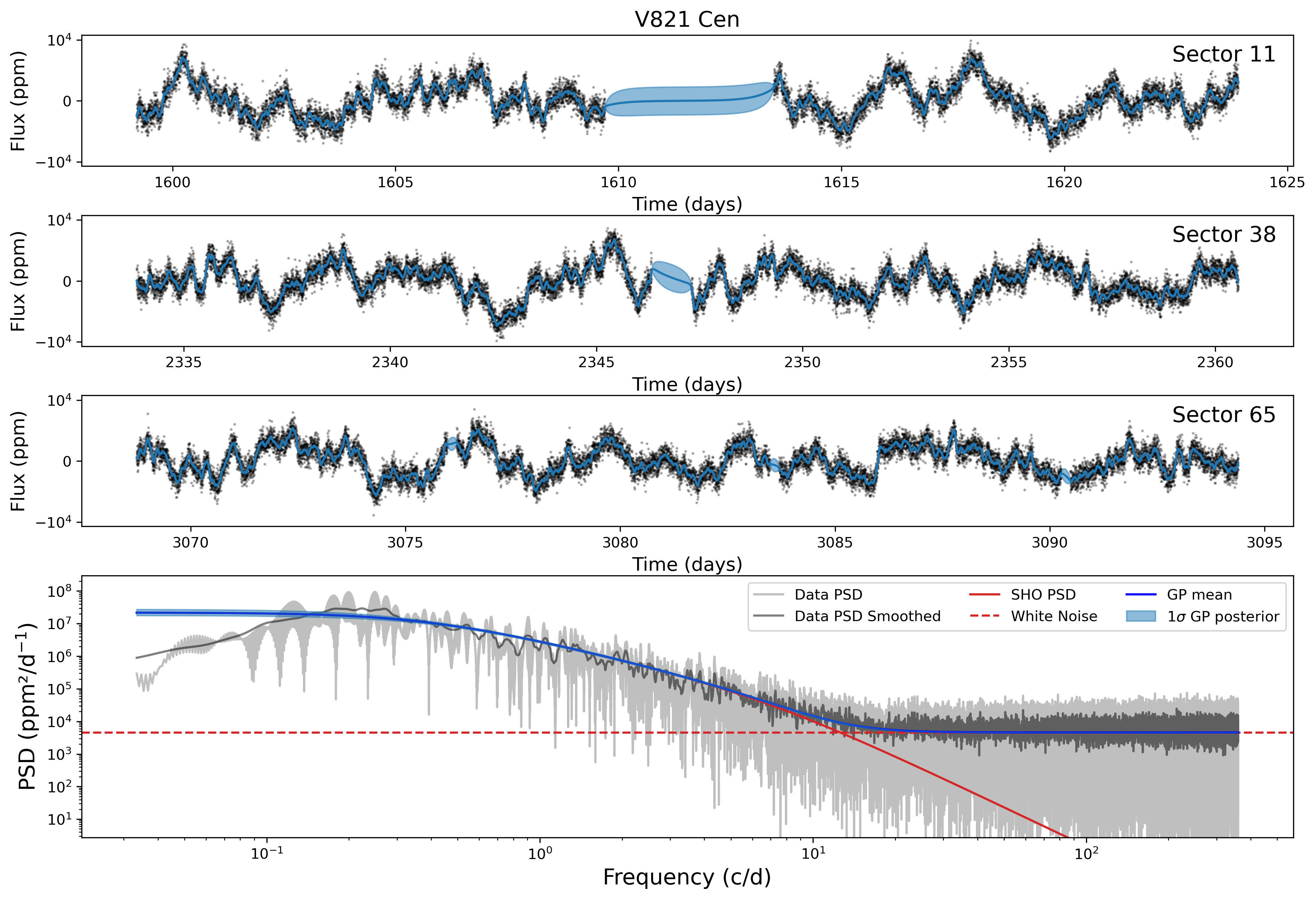}
    \caption{
    Example GPR fit to the \tess{} light curve for a representative target (V821~Cen) and its corresponding PSD. Top panels: Individual sector light curves (black points) shown in ppm, with the GPR fit overplotted (blue line). The blue shaded region indicates the 1$\sigma$ confidence interval. Bottom panel: PSD of the observed light curve (grey), with a smoothed version (dark grey). The dashed red line indicates the white noise level measured by the GP, and the PSD calculated from the GP fit parameters is shown as a solid red line (see Eqn~\ref{eqn:gpr_psd}). The blue curve and shaded region denote the median and 1$\sigma$ interval of the full representative GP PSD, i.e. with white noise added. 
    }
    \label{fig:v821cen_gprfit}
\end{figure*}


\section{Fitting SLF Variability with Gaussian Process Regression}
\label{sec:gpr}

SLF variability is characterised using a wide variety of methods throughout the literature. 
It is most common to fit a Lorentzian-like profile (sometimes called a "Harvey profile") directly to the power spectrum \citep[e.g.][]{Harvey_profile,Kallinger2014_granulation,Bowman2019_slf_psd_fitting,Pedersen2025_slf}. However, one can also use Gaussian Process Regression (GPR) to fit the covariance of the data in the time domain \citep{Bowman2022_slf_GP_MW,Bowman2024_slf_GP_LMC,Zhang2024_RSG_GP_slf} or analyse the cumulative integrated power density for a model-independent method \citep[e.g.][]{Schultz2022_slf_models,Pedersen2025_slf}.
For our analysis, we chose the GPR method because it has an interpretable model and avoids the gap-induced artefacts present in Fourier analyses of multiple \tess{} sectors. For direct comparison with other works, we include Lorentzian-like fits and cumulative integrated power densities in Appendices~\ref{app:harvey}~and~\ref{app:int_power}, respectively.

By applying the GPR method the SLF variability is characterised directly in the time domain, however this method fits the covariance of the data rather than fitting a functional form directly to the data.
A smart choice of a covariance or "kernel" function is central, and a typical physically-motivated choice in stellar variability studies is the stochastically driven/damped simple harmonic oscillator (SHO), implemented in the python packages {\sc celerite} and {\sc celerite2} \citep{celerite1,celerite2}. In \citet{celerite1} Eqn.~20, the power spectral density (PSD) of an SHO process is defined as:
\begin{equation}
\label{eqn:gpr_psd}
    S(\omega) = \sqrt{\frac{2}{\pi}} \frac{S_0 \omega_0^4}{(\omega^2-\omega_0^2)^2 + \omega_0^2 \omega^2/Q^2},
\end{equation}
where $\omega_0$ is the angular frequency characterising the variability, $S_0$ is the power density of the signal at zero frequency, and $Q$ is the quality factor of the oscillation. In an alternate parametrisation of the SHO \citep{celerite2}, $S_0$ is expressed as a characteristic amplitude or standard deviation of the process, $\sigma_A = \sqrt{S_0 \omega_0 Q}$, which corresponds to the rms variability of the time series. As noted in \citet{celerite1} Eqn.~24, in the limit of $Q=1/\sqrt2$, the PSD simplifies to
\begin{equation}
\label{eqn:gpr_psd_simplified}
    S(\omega) = \sqrt{\frac{2}{\pi}} \frac{S_0}{1+(\omega/\omega_0)^4},
\end{equation}
which is identical to the form used in Lorentzian-fitting (see Appendix~\ref{app:harvey}), assuming $\gamma=4$.

In this work, we fitted all EHe \tess{} light curves using GPR, excluding only BX\,Cir and V652\,Her (See Section~\ref{subsubsec:non-linears}). We show a representative fit using V821~Cen as an example in Figure~\ref{fig:v821cen_gprfit}. We used the full light curves and maintained the gaps in between sectors. For BD+37~442 and BD+37~1977, we used the light curves that had been pre-whitened to remove the potential $r$-mode signals (See Section~\ref{subsubsec:rmodes}). For fitting, we used the {\sc celerite2} implementation of a single-term SHO kernel with five free parameters corresponding to a constant mean (consistent with zero), the SHO hyperparameters ($S_0$, $\omega_0$, $Q$), and an additional "jitter" term that is incorporated into the white noise measurement along with the measurement uncertainties reported by TESS. These five parameters are all sampled in logarithmic-space. The initial guesses for the SHO parameters were informed by the observed power density spectrum, with $S_0$ estimated from the low-frequency power, $\omega_0$ initialised to 1.0 d$^{-1}$, and $Q$ set to 0.5, corresponding to the critically damped regime, though our final results do not depend strongly on these initial guesses. Then, we performed an initial fit using a simple maximum-likelihood optimisation from \texttt{scipy}. These fits were then used to define weak Gaussian priors for the MCMC sampling, centred on the maximum-likelihood solution with broad widths ($\sigma$=2 in log-parameter space), ensuring minimal constraint on the posterior. We sampled the posterior distribution of the fitted parameters using the MCMC algorithm {\sc emcee} \citep{emcee} with 32 walkers, 1000 steps for burn-in which are discarded, and 5000 steps for production. We adopted the 16th and 84th percentiles of the posterior as the lower and upper uncertainties for each parameter. All fitted parameters and uncertainties are listed in Table~\ref{tab:gpr_fits}. For the non-detections (see Section~\ref{subsec:weak}), we include our measurements of $S_0$, the white noise level, and the S/N as upper limits.

\begin{landscape}
\begin{table}
    \centering
    \caption{GPR fit parameters}
    \label{tab:gpr_fits}
    \renewcommand{\arraystretch}{1.25} 
    \begin{tabular}{lccccccccc}
        Star & $S_{0 \rm , uncorrected}$ & $S_{0 \rm , corrected}$ & $\omega_0$ & $Q$ & WN & S/N & $\nu_{\rm SLF}$ & $\tau_{\rm SLF}$ & $\tau_{\rm damp}$ \\
         & (ppm$^2$/d$^{-1}$) & (ppm$^2$/d$^{-1}$) & (rad/d) &  & (ppm$^2$/d$^{-1}$) & ($S_0$/WN)& (c/d) & (d) & (d) \\
        \hline
NO Ser & $7.01^{+6.55}_{-2.89}\times 10^{8}$ & $7.79^{+7.28}_{-3.21}\times 10^{8}$ & $1.77^{+0.25}_{-0.27}$ & $0.63^{+0.22}_{-0.15}$ & $6.52\times 10^{3}$ & 107493 & $0.28^{+0.04}_{-0.04}$ & $3.54^{+0.53}_{-0.50}$ & $7.14^{+2.71}_{-2.01}\times 10^{-1}$ \\
V2244 Oph & $3.63^{+4.88}_{-1.70}\times 10^{10}$ & $3.99^{+5.36}_{-1.87}\times 10^{10}$ & $0.75^{+0.13}_{-0.14}$ & $1.24^{+1.03}_{-0.47}$ & $2.79\times 10^{4}$ & 1301039 & $0.12^{+0.02}_{-0.02}$ & $8.38^{+1.58}_{-1.40}$ & $3.32^{+2.82}_{-1.36}$ \\
PV Tel & $1.34^{+0.51}_{-0.33}\times 10^{9}$ & $1.34^{+0.51}_{-0.33}\times 10^{9}$ & $1.45^{+0.10}_{-0.11}$ & $0.91^{+0.17}_{-0.13}$ & $2.72\times 10^{3}$ & 491073 & $0.23^{+0.02}_{-0.02}$ & $4.33^{+0.33}_{-0.31}$ & $1.25^{+0.25}_{-0.21}$ \\
LSS 99 & $5.50^{+1.38}_{-0.99}\times 10^{9}$ & $5.83^{+1.47}_{-1.05}\times 10^{9}$ & $1.10^{+0.05}_{-0.06}$ & $1.78^{+0.42}_{-0.30}$ & $2.46\times 10^{4}$ & 223052 & $0.18^{+0.01}_{-0.01}$ & $5.71^{+0.30}_{-0.28}$ & $3.24^{+0.78}_{-0.56}$ \\
LSS 4357 & $3.34^{+3.99}_{-1.51}\times 10^{10}$ & $5.90^{+7.06}_{-2.67}\times 10^{10}$ & $2.11^{+0.35}_{-0.38}$ & $0.20^{+0.05}_{-0.04}$ & $6.09\times 10^{5}$ & 54826 & $0.34^{+0.06}_{-0.06}$ & $2.98^{+0.54}_{-0.49}$ & $1.86^{+0.54}_{-0.50}\times 10^{-1}$ \\
V1920 Cyg & $4.36^{+0.68}_{-0.54}\times 10^{9}$ & $4.96^{+0.77}_{-0.62}\times 10^{9}$ & $2.07^{+0.07}_{-0.07}$ & $1.40^{+0.16}_{-0.14}$ & $2.93\times 10^{4}$ & 148745 & $0.33^{+0.01}_{-0.01}$ & $3.03^{+0.10}_{-0.10}$ & $1.35^{+0.16}_{-0.14}$ \\
CD-46 11775 & $3.67^{+0.90}_{-0.65}\times 10^{9}$ & $4.43^{+1.09}_{-0.79}\times 10^{9}$ & $2.31^{+0.11}_{-0.11}$ & $1.34^{+0.24}_{-0.19}$ & $7.56\times 10^{4}$ & 48485 & $0.37^{+0.02}_{-0.02}$ & $2.72^{+0.14}_{-0.13}$ & $1.17^{+0.22}_{-0.17}$ \\
EC 19529-4430 & $<1.72^{+0.83}_{-0.53}\times 10^{6}$ &  &  &  & $1.32\times 10^{5}$ & 13 &  &  &  \\
V2205 Oph & $5.42^{+2.06}_{-1.33}\times 10^{9}$ & $5.66^{+2.15}_{-1.39}\times 10^{9}$ & $1.77^{+0.12}_{-0.13}$ & $2.95^{+1.81}_{-0.87}$ & $1.61\times 10^{4}$ & 337254 & $0.28^{+0.02}_{-0.02}$ & $3.55^{+0.26}_{-0.25}$ & $3.33^{+2.07}_{-1.01}$ \\
V5541 Sgr & $7.90^{+2.67}_{-1.79}\times 10^{8}$ & $1.31^{+0.44}_{-0.30}\times 10^{9}$ & $11.91^{+0.89}_{-0.88}$ & $0.47^{+0.07}_{-0.06}$ & $1.63\times 10^{6}$ & 484 & $1.90^{+0.14}_{-0.14}$ & $5.28^{+0.39}_{-0.39}\times 10^{-1}$ & $7.97^{+1.31}_{-1.21}\times 10^{-2}$ \\
V2076 Oph & $1.14^{+0.27}_{-0.20}\times 10^{9}$ & $1.30^{+0.31}_{-0.23}\times 10^{9}$ & $8.77^{+0.43}_{-0.44}$ & $1.14^{+0.19}_{-0.15}$ & $1.17\times 10^{6}$ & 966 & $1.40^{+0.07}_{-0.07}$ & $7.16^{+0.36}_{-0.35}\times 10^{-1}$ & $2.61^{+0.46}_{-0.37}\times 10^{-1}$ \\
BD+37 442 & $1.46^{+0.11}_{-0.10}\times 10^{6}$ & $1.49^{+0.11}_{-0.10}\times 10^{6}$ & $149.36^{+5.52}_{-5.18}$ & $0.25^{+0.01}_{-0.01}$ & $6.21\times 10^{3}$ & 234 & $23.77^{+0.88}_{-0.82}$ & $4.21^{+0.15}_{-0.16}\times 10^{-2}$ & $3.28^{+0.20}_{-0.20}\times 10^{-3}$ \\
BD+37 1977 & $1.95^{+0.17}_{-0.15}\times 10^{6}$ & $1.95^{+0.17}_{-0.15}\times 10^{6}$ & $113.80^{+4.29}_{-4.06}$ & $0.25^{+0.01}_{-0.01}$ & $8.08\times 10^{3}$ & 241 & $18.11^{+0.68}_{-0.65}$ & $5.52^{+0.20}_{-0.21}\times 10^{-2}$ & $4.48^{+0.28}_{-0.28}\times 10^{-3}$ \\
FQ Aqr &   &   &   &   &   &   &   &   &   \\
V4732 Sgr &   &   &   &   &   &   &   &   &   \\
V354 Nor &   &   &   &   &   &   &   &   &   \\
V821 Cen & $2.79^{+0.71}_{-0.53}\times 10^{7}$ & $2.83^{+0.72}_{-0.54}\times 10^{7}$ & $10.04^{+0.57}_{-0.59}$ & $0.23^{+0.02}_{-0.02}$ & $4.56\times 10^{3}$ & 6124 & $1.60^{+0.09}_{-0.09}$ & $6.26^{+0.37}_{-0.36}\times 10^{-1}$ & $4.55^{+0.47}_{-0.45}\times 10^{-2}$ \\
DN Leo & $<6.01^{+2.08}_{-1.43}\times 10^{4}$ &  &  &  & $5.17\times 10^{3}$ & 11 &   &  &  \\
V652 Her &   &   &   &   &   &   &   &   &   \\
BX Cir &   &   &   &   &   &   &   &   &   \\
GALEX J184559.8–413827 &   &   &   &   &   &   &   &   &   \\
EC 20236-5703 &   &   &   &   &   &   &   &   &   \\
LS IV+6 2 & $<8.61^{+2.53}_{-1.42}\times 10^{5}$ &  & &  & $7.22\times 10^{4}$ & 12 & &  &  \\
PG 1415+492 & $<3.82^{+2.05}_{-1.26}\times 10^{7}$ &  &   &   & $5.52\times 10^{6}$ & 7 &   &   &  \\
EC 20111-6902 &   &   &   &   &   &   &   &   &   \\
EC 20262-6000 & $<2.64^{+2.37}_{-1.04}\times 10^{7}$ &  &   &   & $1.27\times 10^{6}$ & 20 &   &   & \\
GALEX J191049.5-441713 & $<4.88^{+1.80}_{-1.19}\times 10^{6}$ &  &   &   & $2.73\times 10^{5}$ & 17 &   &   &  \\
2MASS J18244794-2214291 &   &   &   &   &   &   &   &   &   \\
2MASS J18335703+0529170 &   &   &   &   &   &   &   &   &   \\
DY Cen & $6.56^{+1.14}_{-0.94}\times 10^{9}$ & $9.43^{+1.64}_{-1.35}\times 10^{9}$ & $3.20^{+0.12}_{-0.12}$ & $1.19^{+0.14}_{-0.12}$ & $5.17\times 10^{5}$ & 12683 & $0.51^{+0.02}_{-0.02}$ & $1.97^{+0.07}_{-0.07}$ & $7.46^{+0.93}_{-0.82}\times 10^{-1}$ \\
    \end{tabular}
\end{table}
\end{landscape}

\subsection{GPR Fitting Results}
\label{sec:gpr_results}

\begin{figure}
    \centering
    \includegraphics[width=\columnwidth]{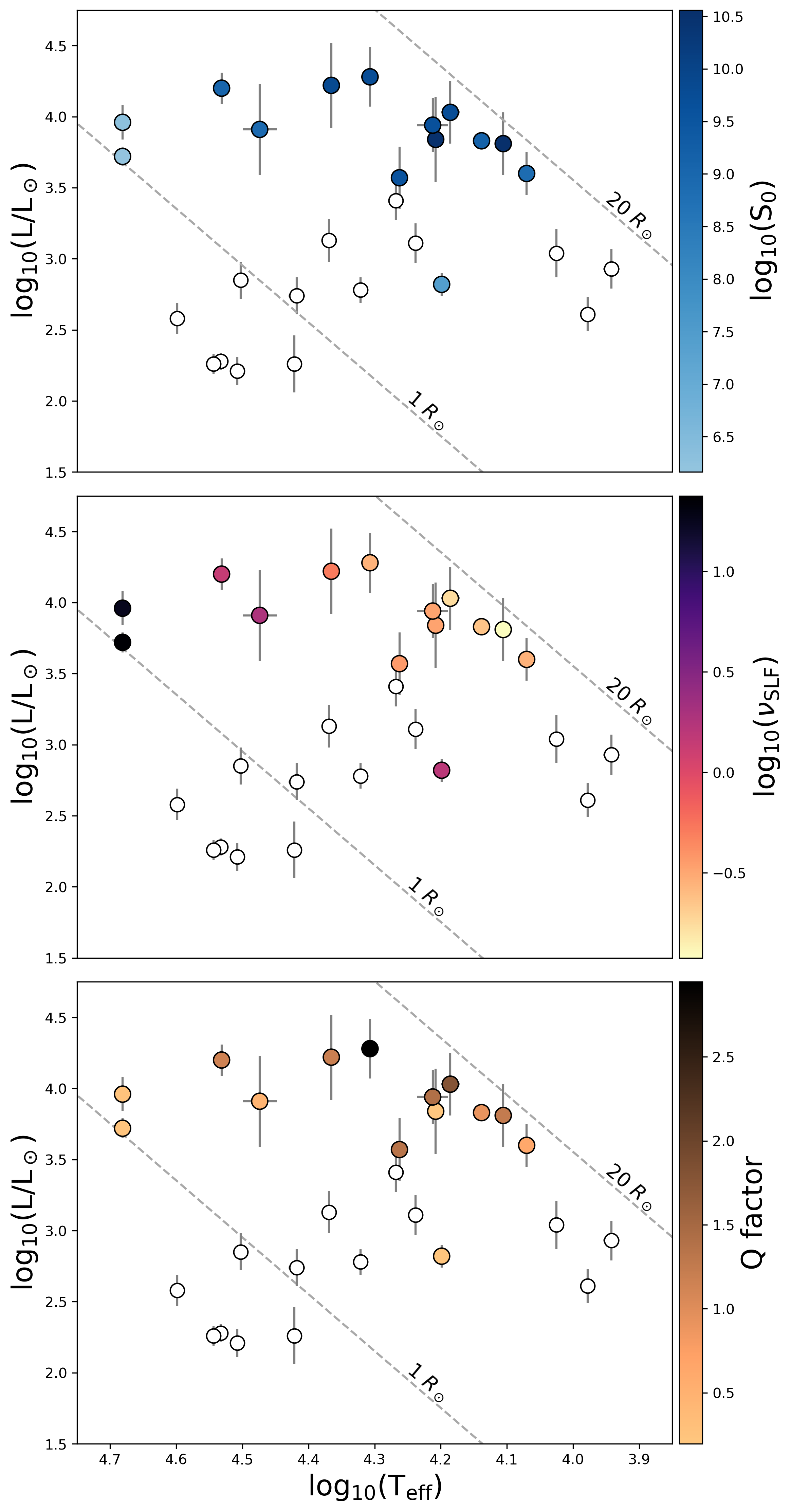}
    \caption{Three figures showing the HRD of known EHe stars with coloured points representing the three parameters fit using GPR as described in the text. Lines of constant radius for 1 $R_\odot$ and 20 $R_\odot$ are plotted as grey dashed lines. Upper: the logarithm of power density at zero frequency in ppm$^2$/d$^{-1}$, $S_0$. Middle: the logarithm of the central frequency of the variability signal $\nu_{\rm SLF}$. Lower: the quality factor of the variability, $Q$. Stars that were not fitted are denoted with white circles.}
    \label{fig:GP_summary}
\end{figure}

We visualise the three SLF variability parameters across the HRD in Figure~\ref{fig:GP_summary}. All three fit parameters vary smoothly with stellar parameters across the range of observed stars. In the upper panel we show $S_0$, which increases towards the upper right of the HRD, as the radius of the stars increases. We note that the lower limit of measured $\log_{10}S_0$ of roughly 5.5 reflects the average white noise levels of the stars, which is $\log_{10}WN\sim$4. 
This amplitude-like scale of the variability increases with stellar radius.
$S_0$ is also the fitted parameter most likely to be affected by crowding in the field, because contamination from nearby stars will dilute the signal of the target. As described in Section~\ref{subsec:contamination}, we have calculated the contamination factors of our targets, which measures the percentage of the flux in the aperture of the light curve that is likely to come from nearby (contaminating) sources. The undiluted $S_0$ can be estimated as $S_0 (1+C)^2$. This does not change the overall trend of $S_0$ with stellar parameters (which varies over orders of magnitude), although some crowded stars may see an increase in $S_0$ by up to a factor of 2. We emphasise that, even after this correction, nearby stars could contribute some of their own SLF variability, especially for high-contamination targets.

In the centre panel of Figure~\ref{fig:GP_summary} we show the parameter $\nu_{\rm SLF} = \omega_0/2\pi$ (sometimes also called $\nu_{\rm char}$), which is the characteristic frequency of the variability. This can be trivially converted to the variability timescale $\tau_{\rm SLF} = 1/\nu_{\rm SLF}$. However, note that the $\nu_{\rm SLF}$ inferred by GPR is not the same as that which would be measured using a Lorentzian-like fitting method, since they fit different functional forms to the data (see e.g. \citealt{Zhang2024_RSG_GP_slf}). In general, $\nu_{\rm SLF,GPR} > \nu_{\rm char,Harvey}$ but there is a large spread in the ratio of the two measurements. In the EHe stars, we find that $\nu_{\rm SLF}$ decreases towards the upper right of the HRD as the radius of the stars increase. 
A further detailed discussion of the variability timescales is provided in context of the possible origin of the signal in Section~\ref{sec:causes}. We can also compare our measured variability timescales ($\tau_{\rm SLF}$) to the pulsation periods measured in past literature for these stars. \citet{jeffery86.hdef.b} measured the period of NO~Ser as 5--8 d, compared to our measurement of 3.53$^{+0.51}_{-0.48}$ d, whereas \citet{morrison87a} measured a period of V2244~Oph of 10--11 d, compared to our 8.34$^{+1.60}_{-1.40}$ d. Our measurements are consistent with both literature values to within 2$\sigma$, which is notable given that the literature estimates assume purely sinusoidal variability. Additionally, literature estimates for the periods of V2205\,Oph and V2076\,Oph are 3--9 d and 0.7--1.1\,d, respectively \citep{jeffery85b,lynasgray87}, compared to 3.55$^{+0.25}_{-0.25}$ d and 0.70$^{+0.03}_{-0.03}$ d. Again, we find close agreement between the literature estimates and our results, despite the simpler pulsation models adopted in previous studies. This supports the robustness of our methodology and motivates future work revisiting ground-based observations with the GPR technique, particularly for stars such as FQ~Aqr, which lack sufficiently long \tess{} baselines to characterise SLF variability.

Finally, in the lowest panel of Figure~\ref{fig:GP_summary} we show how $Q$ varies across stellar parameters. Once again, $Q$ increases as the stellar radius increases. This indicates that the variability becomes "more coherent" as the radius increases, or conversely that stars with smaller radii have increased damping. By definition, $Q=1/2$ indicates a "critically damped" oscillator, and thus $Q<1/2$ indicates an "overdamped" system and $Q>1/2$ indicates an "underdamped" system. From our analysis, it is clear that the variability in EHe stars encompasses both the under- and overdamped cases, and thus a range of damping across the sample.
We calculated the timescale of the damping ($\tau_{\rm damp}$) using the relation 
\begin{equation}
    Q = \frac{\tau_{\rm damp}\omega_0}{2} = \frac{\pi\tau_{\rm damp}}{\tau_{\rm SLF}}.
\end{equation}
We report $\tau_{\rm damp}$ in Table~\ref{tab:gpr_fits}. Thus, stars with lower Q values have shorter damping timescales, or an increased damping rate. It is not clear what physical process in the star causes this damping, however we do see that $\tau_{\rm damp}$ is correlated with $\tau_{\rm SLF}$ (consistent with results from \citealt{Bowman2024_slf_GP_LMC}).


\section{What causes SLF variability in EHe stars?}
\label{sec:causes}

There are four theories currently employed to explain the existence of intrinsic SLF variability in stellar light curves.
We discuss each of these below, as well as a fifth possibility. While instrumental effects can mimic SLF variability, we consider this unlikely given the robustness of the reduction pipelines used in producing \tess{} light curves.

\subsection{Granulation}

\begin{figure}
    \centering
    \includegraphics[width=\columnwidth]{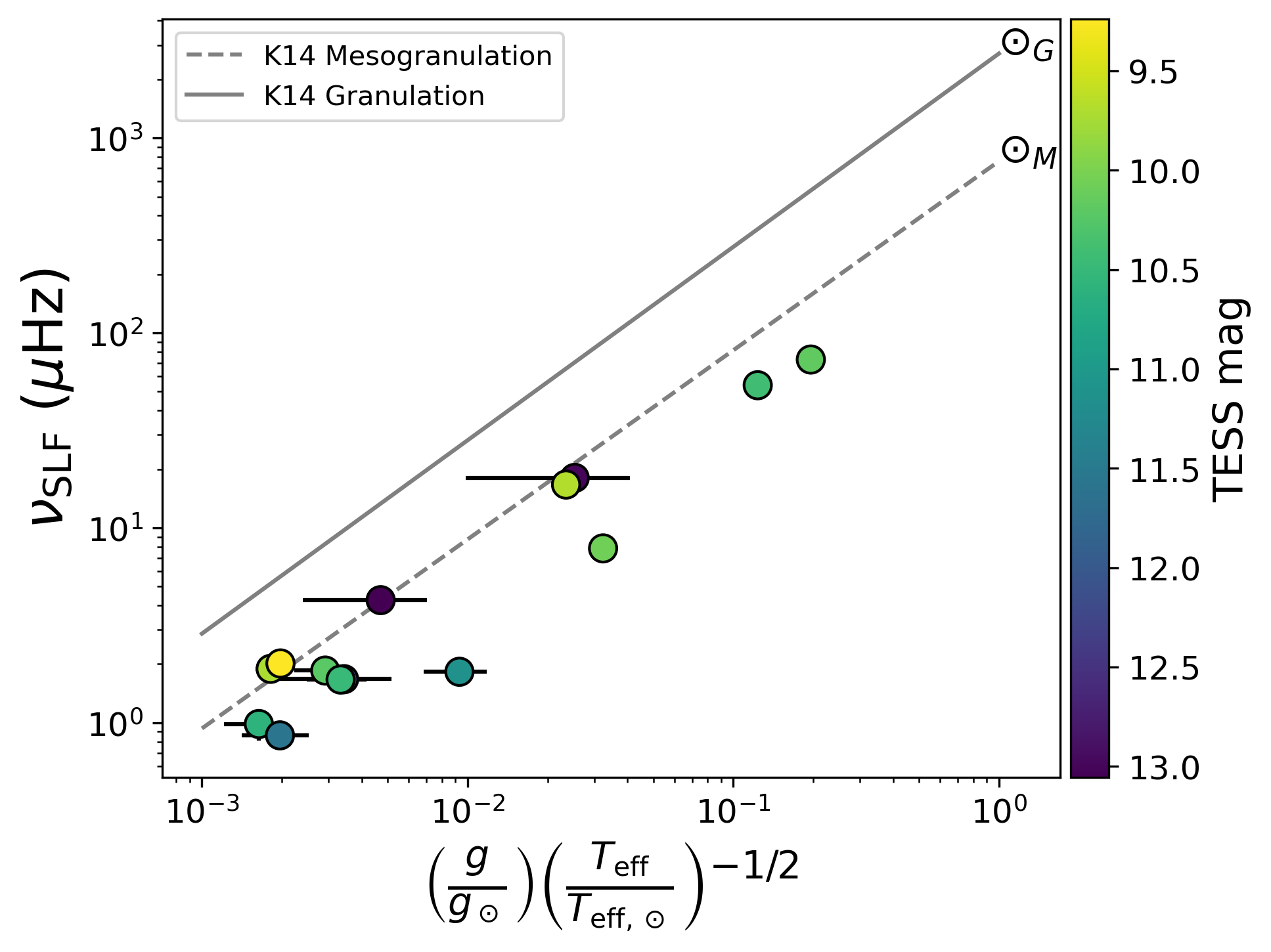}
    \caption{
    Upper: Characteristic SLF variability frequency, $\nu_{\mathrm{SLF}}$, as a function of the scaling $\left(g/g_\odot\right)\left(T_{\mathrm{eff}}/T_{\mathrm{eff},\odot}\right)^{-1/2}$. Filled circles show observational measurements, colour-coded by \tess{} magnitude. The dashed and solid lines indicate the granulation scaling relations ("Harvey-like" components) reported by \citet{Kallinger2014_granulation} for \textit{Kepler} red giants. Note that for this figure only, we use $\nu_{\rm SLF}$ as measured using Harvey profiles (see Appendix~\ref{app:harvey}), for a more direct comparison to the red giant analysis.}
    \label{fig:timescales_gran}
\end{figure}

Granulation describes the photometric variation of a star with a large surface convective layer that has individual convective cells, or granules, that cause a stochastic variability. The prototype case is the Sun \citep[e.g.][]{Harvey_profile}, which shows many different scales of granulation. Variation due to granulation has been observed in cool dwarfs and red giants. Red giant studies have found that the timescale of this stochastic variation scales as $\tau_{\rm gran} \propto L/MT_{\rm eff}^{3.5}$ \citep{Huber2009,KjeldsenBedding2011_amplitudes} which is derived from assuming that a single convective cell moves a distance proportional to one pressure scale height ($H_P$) at a speed proportional to the average sound speed ($c_s$), where $\tau_{\rm gran}\propto H_P/c_s$. 

In Figure~\ref{fig:timescales_gran}, we compare 
\begin{equation}
    \nu_{\rm SLF} (\mu {\rm Hz}) \propto \frac{g}{g_\odot}\left(\frac{T_{\rm eff}}{T_{\rm eff,\odot}}\right)^{-1/2}.
\end{equation}
for the EHe star observations directly with the two timescales of granulation scaling relations measured by \citet{Kallinger2014_granulation} for Kepler red giants. Note that the latter study measured the granulation timescales using Harvey profiles, and we therefore plot $\nu_{\rm SLF}$ for the EHe star observations measured using Harvey profiles for a direct comparison (see Appendix~\ref{app:harvey}). We found that the timescales of EHe SLF variability are roughly comparable to those predicted by the granulation scaling relations, although slightly lower in frequency than the lowest frequency scale of red giant granulation, which is sometimes called mesogranulation. This qualitative agreement is surprising because EHe stars are not expected to have significant surface convective layers or granulation signals.


\subsection{Subsurface Convection}
\label{subsec:fecz}


\begin{figure*}
    \centering
    \includegraphics[width=\textwidth]{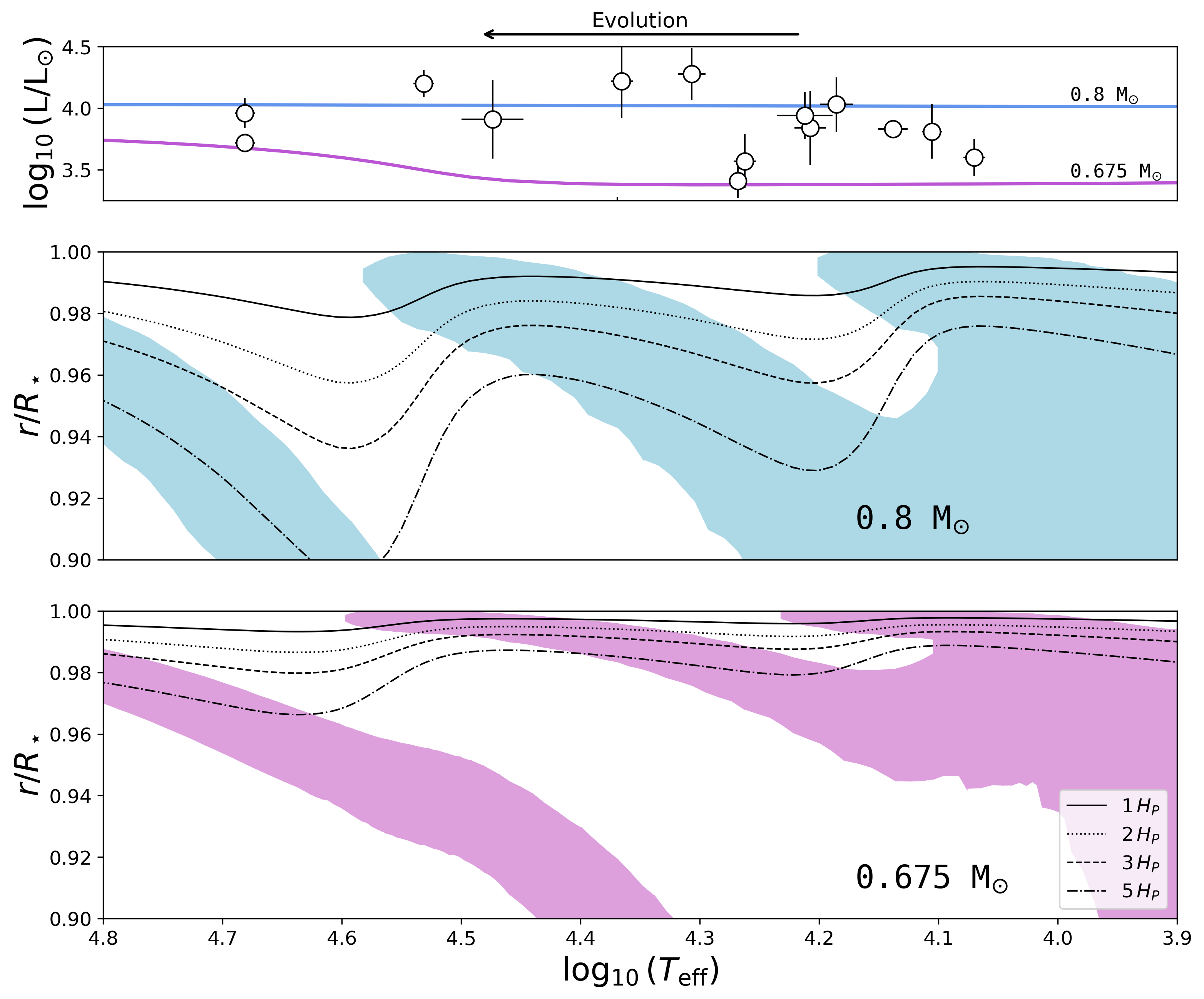}
    \caption{
    Upper: A slice of the Hertzsprung–Russell diagram showing the sample of EHe stars, plotted as open circles. Overlaid are two evolutionary tracks of CO+He merger models \citep{Crawford2024_dlhdcmodels} with differing total mass: 0.8 M$_\odot$ (blue) and 0.675 M$_\odot$ (magenta). The arrow marks the direction of evolution, toward decreasing effective temperature. Centre and Lower: Kippenhahn-style diagrams of the outermost 10\% of the stellar radius for the 0.8 M$_\odot$ and 0.675 M$_\odot$ models shown above, plotted against $\log_{10}(T_{\rm eff})$ to allow direct comparison with the top panel. Blue and magenta shading indicate convective zones in the 0.8 M$_\odot$ and 0.675 M$_\odot$ models respectively. Black curves trace the depth below the stellar surface corresponding to 1, 2, 3, and 5 local pressure scale heights $H_P$, shown as solid, dotted, dashed, and dot-dashed lines respectively.
    }
    \label{fig:kipp}
\end{figure*}

Subsurface convection zones refer to the thin convection zones generated by opacity bumps near the surfaces of one-dimensional massive star models \citep{Cantiello2009_slf,Cantiello2021_slf}. Those models have two convection zones: one induced by the He II partial ionisation opacity peak (the HeCZ), and one induced by the Z-bump or Fe opacity peak (the FeCZ). \citet{Cantiello2021_slf} proposed that SLF variability is driven by the FeCZ rather than the thinner HeCZ, on the basis that the convective properties of the FeCZ are more consistent with those of the observed SLF variability. How the turbulent motions in the FeCZ translate to the surface photometric variability is unclear, but \citet{Cantiello2009_slf} suggested that the FeCZ drives a collection of gravity waves at its upper boundary that travel to the surface. Based on a comparison to 3-D hydrodynamical simulations, \citet{Schultz2022_slf_models} suggested that the existence of the subsurface FeCZ in one-dimensional models is an artefact of the assumptions made in mixing-length theory and that, in three-dimensional models, this FeCZ is able to penetrate through to the surface of the star and present a signal qualitatively similar to the granulation case.

To explore subsurface convection in EHe stars, we examined two of the one-dimensional MESA models of HdC stars computed by \citet{Crawford2024_dlhdcmodels}, with $M_{\rm tot}$ of 0.8 \Msolar{} (${M_{\rm donor}}/{M_{\rm accretor}} = 0.45$) and 0.675 \Msolar{} (${M_{\rm donor}}/{M_{\rm accretor}} = 0.675$). Assuming that EHe stars result from a post-double-white-dwarf merger system, HdC models should evolve into EHe models as they move leftwards across the HRD (as seen in Figure~\ref{fig:ehe_hrd}). We show the convective profiles of two EHe models in Figure~\ref{fig:kipp}. Three thin convective zones, each only a few pressure scale heights deep, migrate outward as the models evolve. These zones are associated with opacity bumps in the models at $\log_{10} T$ of $\sim$4.3--4.5, $\sim$4.5--4.9, and $\sim$5.3, likely due to the effects of He I + C II partial ionisation, He II + C III partial ionisation, and the Z-bump, respectively.

Throughout the EHe evolution, there is consistently at least one subsurface convection zone within five pressure scale heights of the surface, but the specific zone(s) vary with surface temperature. This suggests that subsurface convection may play a role in the observed SLF variability of these stars. This structure also closely parallels the behaviour of the HeCZ and FeCZ in one-dimensional massive star models \citep[e.g.][]{Cantiello2009_slf,Cantiello2021_slf}. To fully explore the impact of these convection zones and their relationship to the variability would require calculations of the convective properties of each of these zones, as well as a range of models with carefully chosen opacity tables and stellar compositions. We leave this detailed analysis to future work.

The FeCZ is closest to the surface on the warmer end of the evolutionary track. Since the Z-bump opacity depends on the Fe abundance, the observed SLF variability could be expected to weaken for stars with lower metallicities. We find two example stars in our sample consistent with this expectation: DN~Leo and EC~19529-4430 (see Figure~\ref{fig:ehe_hrd}). These two stars are considered non-detections of SLF variability in the \tess{} light curves, and are especially metal-weak compared to other EHe stars \citep{heber83,kupfer17,jeffery24}. Two stars with clear SLF variability detections, V821~Cen and CD-46~11775, have similar radii and brightnesses to the two non-detections and are both metal-rich \citep{pandey06a}, reinforcing metal-weakness as a plausible explanation for the non-detections. The models computed by \citet{Crawford2024_dlhdcmodels} adopted a 10\% solar metallicity ($\log\epsilon(\rm Fe)$=6.5, [Fe]=-1) to match observed HdC stars \citep[see e.g.][]{menon19,crawford20}, and this metallicity is similar to the average Fe abundance for the EHe stars \citep{pandey06a,pandey06b,jeffery11a}. Future modelling work should explore even lower metallicities, which may have smaller or less vigorous FeCZs. However, such low metallicities will require a careful treatment of the stellar opacities.

\subsection{IGWs from Core Convection}
\label{subsec:igw}

A popular explanation for SLF variability in massive stars invokes internal gravity waves (IGWs) excited by core convection \citep{Aerts2015_ob_stars}. In this interpretation, core convection excites IGWs at the interface between the core and the radiative envelope. They travel outwards through the thin outer convection zones, and present themselves on the surface of the star as SLF photometric variation. However, EHe stars likely do not have convective cores\footnote{In the merged CO+He case, the cores are carbon-oxygen degenerate. In the merged He+He case, the cores are still helium degenerate, awaiting ignition by an active helium-burning shell.}, so they should not experience IGWs excited by core convection. However, IGWs can also be generated by the thin subsurface convection zones. Recently, \citet{Pathak2026_slf_igw_model} use a 3-D model of a 25 \Msolar{} star to explore the observable photometric variability of the IGWs generated by a thin subsurface convection zone. They interpreted the SLF variability PSD as a "forest" of thin $g$-mode oscillations, which can occasionally exhibit especially large amplitudes. This could lead to the weak oscillation modes in BD+37~1977 and BD+37~442 that have not been definitively identified yet (See Section~\ref{subsubsec:rmodes}). However, it would not explain why those modes exhibit a nearly harmonic structure. Thus, we do not find strong evidence to support the IGW theory in the context of EHe stars.

\subsection{Variable Winds}
\label{subsec:winds}

Using massive star simulations, \citet{Krticka2018_windsSLF,Krticka2021_windsSLF} found that a variable mass-loss rate can create photometric variations that resemble SLF variability. This is due to the variable radiatively-driven winds causing modulated wind blanketing effects that stochastically change the brightness of the star over time. The example simulation used by \citet{Krticka2018_windsSLF} had an average mass-loss rate on the order of $10^{-6}$ \Msolar/yr, which is on the highest end of typical O and B stars (see e.g. \citealt{Krticka2025_masslossrates}). \citet{jeffery10} measure the mass loss rates of EHe stars to be in the range $10^{-10}$--$10^{-7}$ \Msolar/yr. Additionally, they found no evidence for variability in the winds over time, making variable winds an unlikely explanation for EHe star SLF variability. However, there may be a special case for the star DY~Cen, which has produced clumpy dust in the past, making it a ``hot RCB star'' \citep{DeMarco2002_hotRCBs}, although it has not exhibited a visual decline in brightness for over 50 years. This clumpy dust may also indicate variable winds. Polarimetric time-series studies of DY~Cen and other EHe stars in the style of those done for massive stars \citep[e.g.][]{Bailey2024_polarisation} would be useful to explore whether variable winds could be causing EHe star SLF variability. We also note, however, that the wind and dust mechanisms in RCB and EHe stars is not well understood, and therefore may not resemble the radiatively-driven winds in O and B type stars.

\subsection{Highly-damped Oscillation modes}
\label{subsec:modes}

\begin{figure}
    \centering
    \includegraphics[width=\columnwidth]{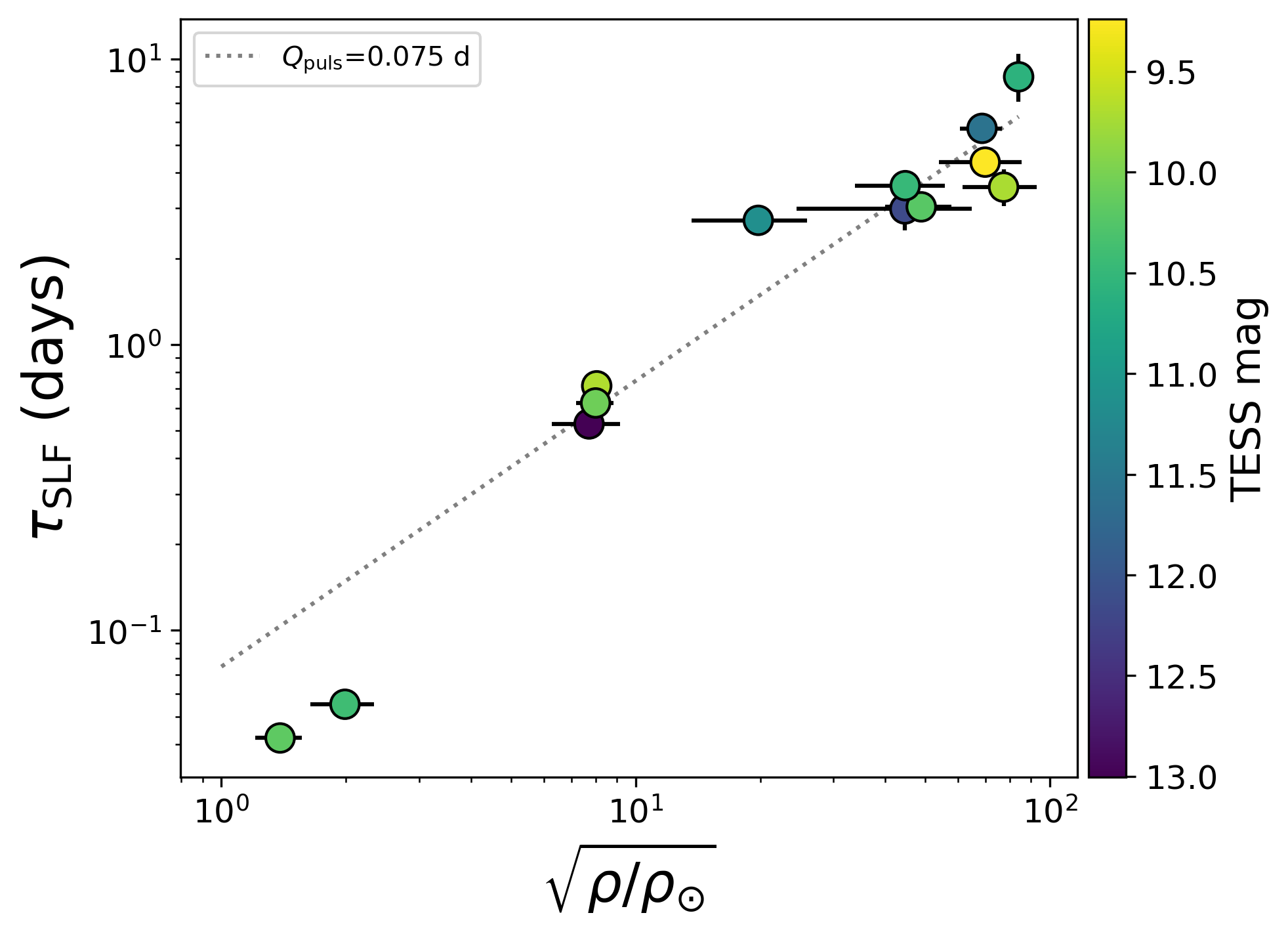}
    \caption{
    SLF variability timescale, $\tau_{\mathrm{SLF}}$, as a function of $\sqrt{\rho/\rho_\odot}$, compared to a timescale predicted for a constant pulsation constant $Q_{\mathrm{puls}} = 0.075$ days (dotted line). Error bars indicate uncertainties on the measured quantities.}
    \label{fig:timescales_puls}
\end{figure}

While not previously discussed as a possible origin of SLF variability, we explore the idea of a near-critically-damped oscillation mode of the star. This type of explanation would lend itself to a straight-forward physical interpretation of the quality factor $Q$ measured for these stars which, as discussed in Section~\ref{sec:gpr_results}, varies with stellar parameters. In fact, the framework of using GPR with an SHO kernel implies that there is a ``true'' excited oscillation frequency that is being highly damped by some stellar process. Historically, many oscillation modes have been suggested as possible origins for EHe stellar variability, such as the strange-modes and non-radial pulsations \citep{jeffery85b,lynasgray87,saio88b,jeffery16a}.

In Figure~\ref{fig:timescales_puls}, we plot $\tau_{\rm SLF}$ versus the mean density of the stars. A linear relationship between these two would be expected if the variability follows a period-mean density relation, or in other words if the variability timescale scales with the dynamical timescale: 
\begin{equation}
    \tau_{\rm SLF} \propto \tau_{\rm dyn} \propto \frac{1}{\sqrt{\rho}} = \sqrt{\frac{R^3}{M}}.
\end{equation}
The data follow a linear scaling reasonably well, and we therefore fitted a line to the data to measure the pulsation constant ($Q_{\rm puls}$\footnote{To aid in clarity, we use $Q_{\rm puls}$ to refer to the pulsation constant, whereas $Q$ is used to refer to the quality factor as fitted by GPR.}). Doing so, we measure $Q_{\rm puls}=0.075$ days, which is a longer timescale than would be expected from the fundamental radial mode ($Q_{\rm puls}\sim0.03$ days). The SLF variability is therefore not consistent with a highly-damped $p$-mode oscillation. However, our analysis does not rule out other pulsation modes such as \textit{strange}-modes as a possible source \citet{saio88b}. More detailed pulsation analysis of EHe-star models within the context of these new measurements would be required to further explore the idea of highly-damped oscillation modes.


\section{Comparisons with other SLF Variables}
\label{sec:slf_comparisons}

SLF variability is common across the HRD--- it is nearly ubiquitous in massive main-sequence stars \citep[e.g.][]{Balona1992_ob_stars,Blomme2011_ob_stars,Tkachenko2014_ob_stars,Aerts2015_ob_stars,Bowman2018_slf_amp_psd_compare,Bowman2019_nature_slf,Bowman2019_slf_psd_fitting,Pedersen2019_ob_stars,Burssens2020_ob_stars,Bowman2020_slf_amp_fitting,Bowman2022_slf_GP_MW,Bowman2024_slf_GP_LMC,Shen2024_ob_stars,Pedersen2025_slf} but also seen in red and yellow supergiants \citep{Kiss2006_RSG_slf,DornWallenstein2019_ysg,Zhang2024_RSG_GP_slf}, blue supergiants \citep{Aerts2018_bsg,Ramiaramanantsoa2018_bsg,Elliott2022_lbv,Ma2024_bsg,Kourniotis2025_bsg}, Wolf-Rayet stars \citep{Lamontange1987_wr_stars,Lepine1999_wr_stars,Moffat2008_wr_stars,chene2011_wr_stars,Ramiaramanantsoa2019_wr_stars,Naze2021_wr_stars,LenoirCraig2022_wr_stars}, and (with a different explanation) in the granulation signals of solar-like oscillators \citep[e.g.][]{KjeldsenBedding2011_amplitudes,Kallinger2014_granulation} and red supergiants \citep{Kiss2006_RSG_slf}. The physical origins of the SLF variability signal are likely different for each class of SLF variable.

Although EHe stars form an exotic stellar class with unusual formation histories, their SLF variability exhibits qualitative similarities to other populations of SLF variables. Across different stellar types, the amplitude of variability (measured by $S_0$) generally increases with luminosity, while the characteristic frequency ($\nu_{\rm SLF}$) decreases with increasing effective temperature (and often luminosity). The quality factor ($Q$) also tends to span a similar range across populations, although absolute values depend on the observational methods and definitions used. These comparisons suggest that, despite differences in formation and mass for EHe stars, their SLF variability is broadly consistent with the stochastic processes seen in other stellar populations, albeit scaled to their characteristic stellar parameters. A full, quantitative comparison across all known SLF variables remains an avenue for future work.

In massive stars, SLF variability exhibits these same general trends in amplitude and frequency. Quantitative comparisons to EHe stars are complicated by differences in luminosity scales. Massive stars are often plotted in a so-called spectroscopic HRD using $\mathcal{L} = T_{\rm eff}^4/g$, equivalent to $L/M$. Since EHe stars have especially high $L/M$ ratios, their position relative to the massive stars differs greatly depending on the HRD representation. Despite this, EHe stars span similar ranges of $Q$, $\nu_{\rm SLF}$, and $S_0$ to massive stars, with a few exceptions in the lowest-amplitude targets. Figure~\ref{fig:timescales_gran} appears in the context of massive stars in articles by \citet{Bowman2019_slf_psd_fitting} (Fig.~8) and \citet{Pedersen2025_slf} (Fig.~8), where they show that the SLF variability timescale for massive stars\footnote{Note that both of these studies used Harvey profiles to measure the SLF variability timescale.} is an order of magnitude lower than predicted by the granulation scaling relations. This likely implies that the driving mechanism for the SLF variability in massive stars is different than that for the EHe stars. 

Red supergiants (RSGs) provide another relevant comparison. Their SLF variability is attributed to granulation in extensive convective envelopes, often accompanied by strong, low-frequency $p$-mode oscillations \citep{Kiss2006_RSG_slf,Goldberg2022_RSG_models,Zhang2024_RSG_GP_slf}. As expected from their larger radii, RSG variability occurs on timescales two to three orders of magnitude longer than those in EHe stars. Amplitude comparisons are challenging due to differing instruments and conventions, but the measured quality factors $Q$ for strongly detected RSGs are broadly consistent with the range seen in EHe stars.


\section{Conclusions \& Outlook}
\label{sec:conclusions}

In the Introduction of this article, we noted seven broad questions that remained about the variability of EHe stars. We summarise our discussion of the \tess{} light curves by revisiting these questions:
\begin{enumerate}
\item \textit{Do all EHes vary in light?} \\
Yes, the majority of EHe stars exhibit variability at some level. We only see non-variability in two cases. Firstly, many of the low-luminosity class of EHe stars (see e.g. \citealt{philipmonai24}) are too faint for \tess{} to measure their variability. We detected variability in only four EHe stars with \tess{} mag $>$ 12: BX\,Cir, a Large Amplitude Pulsator, and three SLF variables, LSS~4357, V5541~Sgr, and DY~Cen. Secondly, we see non-variability in two metal-poor stars: EC~19529-4430 (T=12.021) and DN~Leo (T=10.136). This may indicate that the mechanism driving the SLF variability in EHe stars may be metallicity dependent. 
\item \textit{Are EHe variations periodic or stochastic?} \\
Both types exist in EHe stars, although periodic variations are significantly less common than stochastic variations. We found two types of periodic variables: the large-amplitude pulsators (BX\,Cir and V652\,Her, see e.g. \citealt{Kilkenny2024_bxcir_v652her_tess}) and the potential $r$-mode pulsators (BD+37\,442 and BD+37\,1977, \citealt{jeffery20c}). All other variables, including the potential $r$-mode pulsators, exhibit stochastic variability.
\item \textit{Can EHe variations be associated with a characteristic timescale?} \\
Yes. Despite the stochastic nature of EHe variability, one can measure a characteristic timescale. In practice, with a GPR framework this value represents the underlying oscillation frequency of a highly-damped simple harmonic oscillator. For stars with published estimates of their periods, those are consistent with the SLF variability timescales measured in this work.
\item \textit{If so, are timescales correlated with other properties?} \\
Yes. We see a general trend that stars with larger radii have longer variability timescales (lower frequencies). We see some agreement with predictions of the longest timescales of surface granulation seen in red giant stars. EHe stars do not have thick surface convection zones to create this granulation signal, but one-dimensional models do predict thin subsurface convection zones within a few pressure scale heights of the stellar surface. This may support the findings of 3-D models such as \citet{Schultz2022_slf_models} that the convective velocities of subsurface convection zones in massive stars can extend to the surface of the star. However, we also see that the timescales of the SLF variability in EHe stars scales with the dynamical timescale of the stars. This is perhaps unsurprising, considering $\tau_{\rm dyn}$ and $\tau_{\rm gran}$ depend very similarly on stellar parameters, and can be represented as $\tau_{\rm dyn} \propto R/c_s$ and $\tau_{\rm gran }\propto H_P/c_s$. Regardless, using a period-mean density relation we measure the average $Q_{\rm puls}$ of EHe stars to be 0.075 days, which is more than twice the fundamental radial mode ($Q_{\rm puls}$=0.03 days). Thus, SLF variability in EHe stars is not consistent with a highly-damped $p$-mode oscillation, and any proposed oscillation mode would have to have a period longer than the fundamental $p$-mode.
\item \textit{Are EHe variations global or local?} \\
The stellar variability is global in the sense that it is occurring over the whole star and varies as a function of global stellar parameters, especially the dynamical timescale, but it is likely driven and damped by small-scale local processes such as convection.
\item \textit{What are the driving mechanisms?} \\
The origin of EHe SLF variability is not immediately clear. From our analysis, it seems probable that the variability is related to the near-surface convective properties of the stars. However, further theoretical explorations will be needed to distinguish between different driving mechanisms.
\item \textit{Is there evidence for secular changes?} \\
For the large amplitude pulsators, there is historical evidence for secular changes in pulsation frequencies \citep[e.g.][etc.]{Kilkenny2024_bxcir_v652her_tess}, which is verified by the \tess{} data. For the SLF variables, we do not see evidence for secular changes on the timescale of the \tess{} mission. As more data from this mission becomes available, more robust analyses that take into account possible secular changes may be possible. Additionally, reanalysing published ground-based data for comparison to the modern \tess{} data may provide a better exploration of this question.
\end{enumerate}

This work has been made possible due to the publicly available data of the nearly all-sky \tess{} mission and recent advancements in data analysis techniques such as scalable GPR with physically motivated kernels \citep{celerite1,celerite2}. There remains a wealth of historical ground-based data for the EHe stars, including those where the \tess{} data is insufficient to characterise the variability. In future works, we will apply our GPR methodology to these data to extend the study of EHe stellar variability. Additionally, we will extend this analysis to the related RCB variables and other hydrogen-deficient stars to elucidate a deeper understanding of the potential evolution between these classes.


\section*{Acknowledgements}

We thank the anonymous referee for their constructive feedback on this article. CC and TB gratefully acknowledge support from the Australian Research Council through Laureate Fellowship FL220100117. MGP is the recipient of an Australian Research Council Australian Discovery Early Career Award (project number DE250100146) funded by the Australian Government.

This paper includes data collected with the \textit{\textit{TESS}} mission \citep{TESS}, obtained from the MAST data archive at the Space Telescope Science Institute (STScI). Funding for the \textit{TESS} mission is provided by the NASA Explorer Program and the Science Mission Directorate. STScI is operated by the Association of Universities for Research in Astronomy, Inc., under NASA contract NAS 5-26555. 

This work made use of several publicly available {\tt python} packages: {\tt astropy} \citep{astropy:2013,astropy:2018}, 
{\tt lightkurve} \citep{lightkurve2018},
{\tt matplotlib} \citep{matplotlib2007}, 
{\tt numpy} \citep{numpy2020},
{\tt pandas} \citep{mckinney-proc-scipy-2010_pandas,the_pandas_development_team_2024_13819579}, 
{\tt scipy} \citep{scipy2020} and {\tt astroquery} \citep{astroquery}.

\section*{Data Availability}

All \tess{} light curves used in this article are freely available via MAST. 
All tables presented here can be downloaded via Zenodo.
\typeout{}
\bibliographystyle{mnras}
\bibliography{ehe_tess,ehe}


\appendix

\section{TESS-Localize on BX Cir}
\label{app:tess-localize}

BX\,Cir exhibits spurious peaks in its amplitude spectrum (Figure~\ref{fig:bxcir}), which we attribute to contaminating sources. To identify their origin, we employed the publicly available \texttt{tess\_localize} package \citep{TESS-localize}, which enables localisation of periodic signals at the pixel level within \tess{} Target Pixel Files (TPFs). \texttt{tess\_localize} operates by taking a user-specified list of frequencies and, for each pixel in the TPF, fitting a sinusoidal model at those frequencies in the time domain. The resulting best-fit amplitudes are then mapped across the pixel grid, producing a spatial amplitude distribution for each frequency. For a genuine astrophysical signal, this distribution approximates the \tess{} point spread function centred on the pixel location of the source, allowing the origin of the signal to be localised. We applied this procedure using the set of frequencies identified via prewhitening (Table~\ref{tab:bxcir}). The resulting amplitude maps revealed that the signals at $2f_{\rm EB2}$ and $2f_{\rm EB1}$ do not originate from the target pixel of BX\,Cir, but instead from two spatially distinct locations in the TPF.

\begin{figure}
    \centering
    \includegraphics[width=\columnwidth]{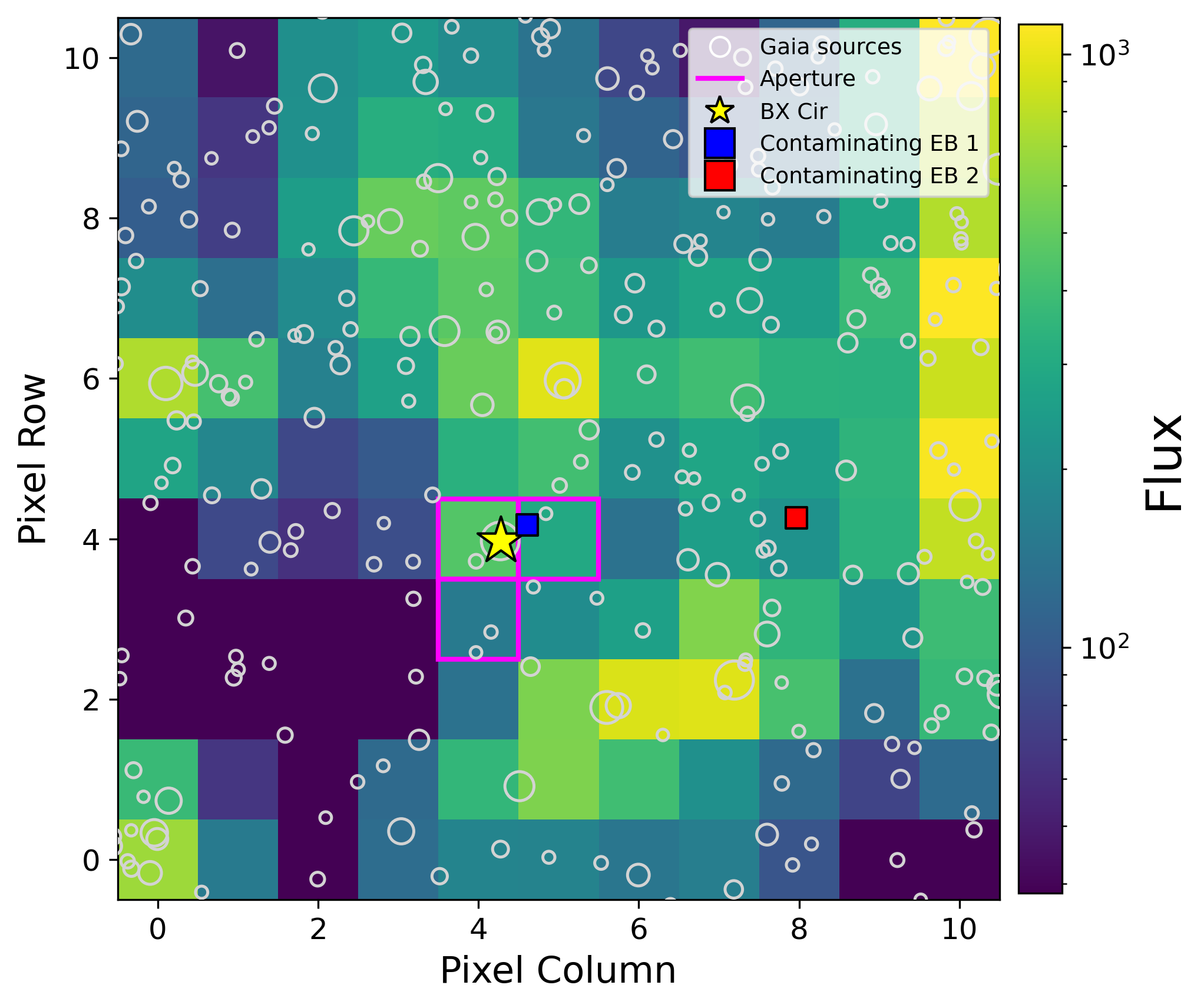}
    \caption{TESS TPF of BX\,Cir from Sector 65. The aperture used for the SPOC light curve is outlined in magenta. We denote BX\,Cir with a yellow star, and the two contaminating eclipsing binaries in blue (Gaia~DR3~5851089030362801152, labelled EB 1) and red (Gaia~DR3~5851089339600429824, labelled EB 2) squares. The Gaia sources with G$<$18 mag are labelled with white circles, where the size of the circle indicates the brightness of the target with smaller circles meaning a fainter star.}
    \label{fig:bxcir_tpf}
\end{figure}

To associate these localised signals with astrophysical sources, we queried catalogued objects in the vicinity of the identified pixel positions using {\sc SIMBAD}. In both cases, the positions coincide with known eclipsing binaries Gaia DR3 5851089030362801152 (EB1, G=17.024) and Gaia DR3 5851089339600429824 (EB2, G=17.215) \citep{GaiaDR3_EBcatalog}. Furthermore, the catalogued orbital periods of these systems are consistent with the frequencies (and harmonics) detected in the BX\,Cir amplitude spectrum. Figure~\ref{fig:bxcir_tpf} shows the Sector 65 representative TPF for BX\,Cir, with the locations of the two contaminating systems indicated in blue and red, respectively.

\section{Harvey Profile Fitting}
\label{app:harvey}

In addition to the GPR characterisation in the main text, we include fits to the PSD of our stars using a Lorentzian-like (``Harvey''; \citealt{Harvey_profile}) profile. Specifically, we fit a function of the form
\begin{equation}
\label{eqn:harvey}
    P(\nu) = \frac{\eta(\nu)a_0}{1+(\frac{\nu}{\nu_{\rm char}})^\gamma} + C_{\rm white}
\end{equation}
where $a_0$ indicates the power density at zero frequency, $\nu_{\rm char}$ indicates the characteristic frequency or ``knee'' in the power spectrum, $\gamma$ represents the ``slope'' of the variability in log-log space, and $C_{\rm white}$ is a constant white noise term. The factor
\begin{equation}
    \eta(\nu) = {\rm sinc}^2 \left( \frac{\pi}{2} \frac{\nu}{\nu_{\rm Nyquist}} \right).
\end{equation}
corrects for the attenuation of the signal towards the Nyquist frequency.

Because $\eta(\nu)$ depends on the cadence and to mitigate windowing effects in Fourier space, each \tess{} sector is treated separately with its own Nyquist frequency and corresponding PSD. For a given star, we construct a joint likelihood over all available sectors by summing the log-likelihood contributions from each sector's Fourier-domain, where each sector likelihood is assumed to follow a $\chi^2$ distribution with two degrees of freedom. The fitted parameters $a_0$, $\nu_{\rm char}$, and $\gamma$ are shared across all sectors, while the white noise term $C_{\rm white}$ is allowed to vary independently between sectors to account for differing noise levels. We sample the posterior distribution using {\sc emcee} with uninformative priors \citep{emcee}. We report our measured values in Table~\ref{tab:harvey_fits}.

\begin{table}
    \centering
    \caption{Harvey fit parameters}
    \label{tab:harvey_fits}
    \renewcommand{\arraystretch}{1.25} 
    \begin{tabular}{lccr}
        Star & $S_{0 \rm , uncorrected}$ & $\nu_{\rm SLF}$ & $\gamma$ \\
         & (ppm$^2$/d$^{-1}$) & (c/d) &   \\
        \hline
NO Ser & $6.45^{+1.62}_{-1.13}\times 10^{8}$ & $0.16^{+0.02}_{-0.02}$ & $2.33^{+0.02}_{-0.02}$ \\
V2244 Oph & $3.31^{+1.73}_{-0.91}\times 10^{10}$ & $0.09^{+0.01}_{-0.02}$ & $2.06^{+0.01}_{-0.01}$ \\
PV Tel & $1.26^{+0.17}_{-0.13}\times 10^{9}$ & $0.17^{+0.01}_{-0.01}$ & $2.78^{+0.02}_{-0.02}$ \\
LSS 99 & $2.14^{+0.43}_{-0.33}\times 10^{10}$ & $0.07^{+0.01}_{-0.01}$ & $2.27^{+0.01}_{-0.01}$ \\
LSS 4357 & $1.25^{+0.20}_{-0.16}\times 10^{10}$ & $0.14^{+0.01}_{-0.01}$ & $2.59^{+0.02}_{-0.02}$ \\
V1920 Cyg & $1.03^{+0.08}_{-0.07}\times 10^{10}$ & $0.16^{+0.01}_{-0.01}$ & $2.37^{+0.01}_{-0.01}$ \\
CD-46 11775 & $9.21^{+1.30}_{-1.01}\times 10^{9}$ & $0.16^{+0.01}_{-0.01}$ & $2.34^{+0.01}_{-0.01}$ \\
EC 19529-4430 &   &   &   \\
V2205 Oph & $3.22^{+0.94}_{-0.63}\times 10^{10}$ & $0.14^{+0.01}_{-0.01}$ & $2.80^{+0.03}_{-0.03}$ \\
V5541 Sgr & $4.10^{+0.27}_{-0.26}\times 10^{8}$ & $1.56^{+0.06}_{-0.06}$ & $3.15^{+0.06}_{-0.06}$ \\
V2076 Oph & $1.18^{+0.07}_{-0.07}\times 10^{9}$ & $1.44^{+0.03}_{-0.04}$ & $3.50^{+0.02}_{-0.02}$ \\
BD+37 442 & $1.00^{+0.03}_{-0.02}\times 10^{6}$ & $6.31^{+0.13}_{-0.13}$ & $2.11^{+0.01}_{-0.01}$ \\
BD+37 1977 & $1.42^{+0.04}_{-0.04}\times 10^{6}$ & $4.66^{+0.11}_{-0.12}$ & $2.06^{+0.01}_{-0.01}$ \\
FQ Aqr &   &   &   \\
V4732 Sgr &   &   &   \\
V354 Nor &   &   &   \\
V821 Cen & $1.34^{+0.07}_{-0.07}\times 10^{7}$ & $0.68^{+0.02}_{-0.02}$ & $2.65^{+0.02}_{-0.02}$ \\
DN Leo &   &   &   \\
V652 Her &   &   &   \\
BX Cir &   &   &   \\
GALEX J184559.8–413827 &   &   &   \\
EC 20236-5703 &   &   &   \\
LS IV+6 2 &   &   &   \\
PG 1415+492 &   &   &   \\
EC 20111-6902 &   &   &   \\
EC 20262-6000 &   &   &   \\
GALEX J191049.5-441713 &   &   &   \\
2MASS J18244794-2214291 &   &   &   \\
2MASS J18335703+0529170 &   &   &   \\
DY Cen & $9.46^{+0.56}_{-0.54}\times 10^{9}$ & $0.37^{+0.01}_{-0.01}$ & $2.91^{+0.02}_{-0.02}$ \\
    \end{tabular}
\end{table}

\section{Cumulative Integrated Power Density}
\label{app:int_power}

As a non-parametric description of the SLF variability, we compute the cumulative integrated power from the PSD and the root-mean-squared (RMS) variability of the light curve, following the methodology of \citet{Pedersen2025_slf}. This approach does not rely on fitting a specific functional form, but instead characterises how power is distributed as a function of frequency. We begin by constructing the cumulative, normalised integrated power of the PSD,
\begin{equation}
P_{\rm int}(\nu) = \frac{\int_{\nu_0}^{\nu} S(\nu'),{\rm d}\nu'}{\int_{\nu_0}^{\nu_{\rm norm}} S(\nu'),{\rm d}\nu'},
\end{equation}
where $S(\nu)$ is the PSD. Here, $\nu_0$ defines the lower frequency bound of the analysis, while $\nu_{\rm norm}$ is the upper frequency at which the cumulative power is normalised. We adopt $\nu_0 = 0.1$ c/d as the lower frequency limit. For consistency across all targets and sectors, we fix $\nu_{\rm norm}$ to the Nyquist frequency corresponding to the longest cadence data used in this work (30-min cadence, $f_{\rm Nyq}=24$ d$^{-1}$), ensuring that the normalisation is applied over a common frequency range.

From $P_{\rm int}(\nu)$, we determine the characteristic frequencies $\nu_{20}$, $\nu_{50}$, and $\nu_{80}$, defined such that
\begin{equation}
P_{\rm int}(\nu_{20}) = 0.2, \quad
P_{\rm int}(\nu_{50}) = 0.5, \quad
P_{\rm int}(\nu_{80}) = 0.8.
\end{equation}
These correspond to the frequencies below which 20\%, 50\%, and 80\% of the total power is contained, respectively. We also compute a dimensionless width parameter,
\begin{equation}
w = \frac{\nu_{80} - \nu_{20}}{\nu_{50}},
\end{equation}
which provides a measure of the approximate slope of the power distribution. The resulting values of RMS, $\nu_{20}$, $\nu_{50}$, $\nu_{80}$, and $w$ for each star are reported in Table~\ref{tab:intpower}. If the star was observed for multiple sectors, we report the average value and the uncertainties reflect the difference between the average with the minimum value (lower uncertainty) and the upper value (upper uncertainty). If the star was observed for only one sector, we do not report an uncertainty.

\begin{table*}
    \centering
    \caption{Cumulative Integrated Power Density Parameters}
    \label{tab:intpower}
    \renewcommand{\arraystretch}{1.25} 
    \begin{tabular}{lccccr}
        Star & RMS & $\nu_{\rm 20}$  & $\nu_{\rm 50}$ & $\nu_{\rm 80}$ & $w$ \\
         & ppm & (c/d) & (c/d) & (c/d) &   \\
        \hline
NO Ser & $9.43\times 10^{3}$ & $0.15$ & $0.25$ & $0.46$ & $1.22$ \\
V2244 Oph & $4.54\times 10^{4}$ & $0.15$ & $0.17$ & $0.50$ & $2.02$ \\
PV Tel & $1.19^{+0.25}_{-0.31}\times 10^{4}$ & $0.16^{+0.02}_{-0.02}$ & $0.23^{+0.03}_{-0.02}$ & $0.79^{+0.60}_{-0.37}$ & $2.58^{+2.22}_{-1.35}$ \\
LSS 99 & $3.54^{+0.78}_{-0.47}\times 10^{4}$ & $0.14^{+0.02}_{-0.02}$ & $0.18^{+0.00}_{-0.00}$ & $0.26^{+0.06}_{-0.05}$ & $0.64^{+0.26}_{-0.35}$ \\
LSS 4357 & $3.49\times 10^{4}$ & $0.15$ & $0.25$ & $0.43$ & $1.13$ \\
V1920 Cyg & $4.03^{+0.59}_{-0.37}\times 10^{4}$ & $0.17^{+0.02}_{-0.03}$ & $0.28^{+0.08}_{-0.07}$ & $0.49^{+0.07}_{-0.09}$ & $1.19^{+0.31}_{-0.14}$ \\
CD-46 11775 & $3.61^{+0.48}_{-0.78}\times 10^{4}$ & $0.19^{+0.01}_{-0.03}$ & $0.33^{+0.05}_{-0.04}$ & $0.70^{+0.20}_{-0.12}$ & $1.59^{+0.91}_{-0.61}$ \\
EC 19529-4430 &   &   &   &   &   \\
V2205 Oph & $5.84\times 10^{4}$ & $0.22$ & $0.27$ & $0.39$ & $0.63$ \\
V5541 Sgr & $2.75\times 10^{4}$ & $0.33$ & $0.76$ & $2.37$ & $2.70$ \\
V2076 Oph & $4.34\times 10^{4}$ & $0.48$ & $0.90$ & $1.85$ & $1.53$ \\
BD+37 442 & $2.86^{+0.02}_{-0.02}\times 10^{3}$ & $1.61^{+0.14}_{-0.14}$ & $4.16^{+0.43}_{-0.43}$ & $10.10^{+0.34}_{-0.34}$ & $2.06^{+0.17}_{-0.17}$ \\
BD+37 1977 & $3.01^{+0.10}_{-0.10}\times 10^{3}$ & $0.90^{+0.22}_{-0.22}$ & $2.89^{+0.28}_{-0.28}$ & $9.05^{+0.52}_{-0.52}$ & $2.87^{+0.54}_{-0.54}$ \\
FQ Aqr &   &   &   &   &   \\
V4732 Sgr &   &   &   &   &   \\
V354 Nor &   &   &   &   &   \\
V821 Cen & $3.31^{+0.20}_{-0.27}\times 10^{3}$ & $0.21^{+0.04}_{-0.03}$ & $0.37^{+0.05}_{-0.03}$ & $1.07^{+0.12}_{-0.12}$ & $2.38^{+0.43}_{-0.55}$ \\
DN Leo & $4.62^{+0.13}_{-0.13}\times 10^{2}$ & $1.28^{+0.19}_{-0.19}$ & $8.85^{+0.10}_{-0.10}$ & $17.53^{+0.16}_{-0.16}$ & $1.84^{+0.02}_{-0.02}$ \\
V652 Her &   &   &   &   &   \\
BX Cir &   &   &   &   &   \\
GALEX J184559.8–413827 &   &   &   &   &   \\
EC 20236-5703 &   &   &   &   &   \\
LS IV+6 2 & $1.71\times 10^{3}$ & $0.80$ & $6.94$ & $17.58$ & $2.42$ \\
PG 1415+492 & $8.37^{+0.57}_{-0.57}\times 10^{3}$ & $2.08^{+0.75}_{-0.75}$ & $8.76^{+1.03}_{-1.03}$ & $18.09^{+0.71}_{-0.71}$ & $1.85^{+0.22}_{-0.22}$ \\
EC 20111-6902 &   &   &   &   &   \\
EC 20262-6000 & $6.55^{+1.16}_{-0.78}\times 10^{3}$ & $3.24^{+1.16}_{-1.75}$ & $10.74^{+1.16}_{-1.34}$ & $18.60^{+0.29}_{-0.50}$ & $1.45^{+0.31}_{-0.24}$ \\
GALEX J191049.5-441713 & $3.16^{+0.33}_{-0.25}\times 10^{3}$ & $2.46^{+1.20}_{-1.60}$ & $9.62^{+1.94}_{-3.07}$ & $17.88^{+0.97}_{-1.42}$ & $1.70^{+0.69}_{-0.40}$ \\
2MASS J18244794-2214291 &   &   &   &   &   \\
2MASS J18335703+0529170 &   &   &   &   &   \\
DY Cen & $5.70^{+1.05}_{-0.96}\times 10^{4}$ & $0.24^{+0.05}_{-0.07}$ & $0.47^{+0.03}_{-0.04}$ & $0.74^{+0.04}_{-0.06}$ & $1.07^{+0.17}_{-0.10}$ \\
    \end{tabular}
\end{table*}

\section{Additional Figures}
\label{app:add_figs}



\bsp	
\label{lastpage}
\end{document}